\documentclass[superscriptaddress,aps,prd,nofootinbib,preprintnumbers]{revtex4-2}

\usepackage{graphicx}
\usepackage{dcolumn}
\usepackage{hyperref}
\usepackage{ulem}
\usepackage{color}
\usepackage{bm}
\usepackage{siunitx}
\usepackage{braket}
\usepackage{amsfonts,amsmath,amssymb,bm,tensor}
\usepackage{ifpdf}
\usepackage{slashed}
\usepackage[mathscr]{eucal}
\usepackage[utf8]{inputenc}
\usepackage{cancel}
\usepackage{enumerate}
\usepackage{mathtools}
\usepackage{physics}
\usepackage{multirow}
\usepackage{tikz}
\usepackage{tikz-3dplot}
\usepackage[compat=1.1.0]{tikz-feynhand}
\usepackage{subfigure}
\begin{document}

%%%%%%%%%%%%%%%%%% Title %%%%%%%%%%%%%%%%%%

\title{Universality of linear perturbations in SU(\texorpdfstring{$N$}{})-natural inflation}
\author{Tomohiro Fujita}
\affiliation{Waseda Institute for Advanced Study, Waseda University, Shinjuku, Tokyo 169-8050, Japan}
\affiliation{Research Center for the Early Universe, University of Tokyo, Bunkyo, Tokyo 113-0033, Japan}

\author{Kai Murai}
\affiliation{ICRR, University of Tokyo, Kashiwa, 277-8582, Japan}
\affiliation{Kavli IPMU (WPI), UTIAS, University of Tokyo, Kashiwa, 277-8583, Japan}

\author{Ryo Namba}
\affiliation{RIKEN Interdisciplinary Theoretical and Mathematical Sciences (iTHEMS), Wako, Saitama 351-0198, Japan}

\begin{abstract}
	We prove the universality of predictions for linear perturbations from the entire class of models of inflation driven by a pseudo-scalar field coupled to an SU($N$) gauge boson, where SU($2$) subgroups in the SU($N$) crossed with the background spatial SO($3$) spontaneously break into a single SO($3$). 
	The effect of which SU($2$) subgroup in SU($N$) acquires a VEV through spontaneous symmetry breaking can be quantified by a single parameter $\lambda$, which always appears in combination with the gauge coupling constant $g$.
	In the linear perturbations, as well as the background system, the same dynamics and predictions as in the chromo-natural inflation hold for its SU($N$) extension by replacing $g \to g\lambda$. 
	The latter class of models thereby draws the same prediction curve on the $n_s$--$r$ plane as the former at the tree level as long as $g \lambda$ stays constant during inflation. We briefly discuss possible transitions from one value of $\lambda$ to another during inflation and the observational prospects.
\end{abstract}

\preprint{RIKEN-iTHEMS-Report-22}

\maketitle

\tableofcontents

%%%%%%%%%%%%%%%%%%%%%%%%%%%%%%%%%%%%%%%%%%%%
\section{Introduction}
\label{sec: intro}
%%%%%%%%%%%%%%%%%%%%%%%%%%%%%%%%%%%%%%%%%%%%

Cosmic inflation successfully addresses the flatness and horizon problems in the big bang cosmology and explain seed fluctuations of the cosmic microwave background (CMB) anisotropies.
Inflation is driven by some field(s) called inflaton and the success of inflation relies on the flatness of the inflaton potential.
In order to keep the flatness of the potential in the presence of radiative corrections, the potential parameter must be fine-tuned,
unless some additional mechanism is at work.
If axion-like particle (ALP) acts as the inflaton, called natural inflation, its shift symmetry protects the potential from radiative corrections and then the flat potential can be naturally realized~\cite{Freese:1990rb,Adams:1992bn,Kim:2004rp}.
However, the natural inflation requires an axion decay constant as large as the Planck scale~\cite{Freese:2004un}, which typically suggests a global symmetry associated with the axion be broken above the quantum gravity scale \cite{Arkani-Hamed:2003xts} and is considered difficult to realize in string theory~\cite{Banks:2003sx}.

Soon after the realization of this issue, a number of mechanisms have been proposed to evade the difficulty and effectively achieve natural inflation, such as by invoking two \cite{Kim:2004rp} or more \cite{Dimopoulos:2005ac,Easther:2005zr} axion fields or $4$-form fields \cite{Kaloper:2008fb}, by the help of extra dimensions \cite{Arkani-Hamed:2003xts}, by exploiting non-periodic contributions to the axion potential called monodromy \cite{Silverstein:2008sg,McAllister:2008hb,Flauger:2009ab}, or by increased friction due to dissipation to other fields \cite{Anber:2009ua}.
The simplest models of axion inflation with cosine-type and monodromy potentials are marginally outside of the $2\sigma$ range of the CMB data \cite{Planck:2018jri}.
Axions naturally couple to gauge fields, and the phenomenology of this coupling in cosmological settings has been extensively studied, such as inflationary model buildings \cite{Anber:2009ua,Peloso:2015dsa,Notari:2016npn,Ferreira:2017lnd,Tangarife:2017vnd,Tangarife:2017rgl,Almeida:2018pir}, CMB observables \cite{Lue:1998mq,Barnaby:2010vf,Sorbo:2011rz,Barnaby:2011vw,Barnaby:2011pe,Dimopoulos:2012av,Anber:2012du,Meerburg:2012id,Linde:2012bt,Ferreira:2014zia,Bartolo:2015dga,Ferreira:2015omg,Peloso:2016gqs,Alexander:2017bxe,Domcke:2018eki,Almeida:2019hhx,Domcke:2019qmm,Domcke:2020zez}, generation of magnetic fields \cite{Durrer:2010mq,Ng:2015ewp,Fujita:2015iga,Adshead:2016iae,Caprini:2017vnn,Shtanov:2019civ,Shtanov:2019gpx,Patel:2019isj,Fujita:2019pmi,Sobol:2019xls} and baryon asymmetry \cite{Jimenez:2017cdr}, formation of primordial black holes \cite{Bugaev:2013fya,Erfani:2015rqv,Domcke:2017fix,Cheng:2018yyr,Ozsoy:2020kat}, dark matter physics \cite{Kamada:2017cpk,Agrawal:2017eqm,Co:2018lka,Bastero-Gil:2018uel,Agrawal:2018vin,Machado:2018nqk}, and sources of gravitational wave \cite{Barnaby:2012xt,Cook:2013xea,Shiraishi:2013kxa,Mukohyama:2014gba,Mirbabayi:2014jqa,Namba:2015gja,Domcke:2016bkh,Shiraishi:2016yun,Obata:2016oym,Ozsoy:2020ccy,Cook:2011hg,Barnaby:2011qe,Crowder:2012ik,Garcia-Bellido:2016dkw,Obata:2016tmo,Obata:2016xcr,Machado:2019xuc,Okano:2020uyr,Shiraishi:2016yun,Bartolo:2016ami,Co:2021rhi}, while some of the models have been directly tested by the Planck mission \cite{Ade:2013ydc,Ade:2015lrj,Ade:2015ava,Akrami:2019izv}.

As another possibility, interactions between  a pseudo-scalar inflaton and non-Abelian gauge fields can induce the slow-roll inflation even with a sub-Planckian decay constant.
In particular, the model where the pseudo-scalar inflaton is coupled with SU(2) gauge fields through the Chern--Simons coupling, Chromo-Natural Inflation (CNI)~\cite{Adshead:2012kp}, has attracted much attention.%
\footnote{In fact, the first introduction of non-Abelian gauge fields in the inflationary contexts is in the so-called gauge-flation model, in which an SU($2$) gauge field alone drives inflation \cite{Maleknejad:2011jw,Maleknejad:2011sq}, which can be viewed as a massive axion limit of CNI \cite{Adshead:2012qe} and which is unfortunately disfavored by observations \cite{Namba:2013kia}.}
In this model, the inflaton velocity induces a homogeneous, isotropic, and attractor solution of the gauge fields~\cite{Maleknejad:2013npa,Wolfson:2020fqz,Wolfson:2021fya}, while the gauge field background slows down the motion of the inflaton.
Since the gauge field background spontaneously breaks the spatial SO(3) rotation symmetry and SU(2) gauge symmetry into the diagonal SO(3) symmetry, the gauge fields can be regarded as a second-order tensor.
Interestingly, one polarization of the tensor components of the gauge field perturbations experiences a tachyonic instability and then significantly sources chiral gravitational waves~\cite{Dimastrogiovanni:2012ew}.
As a result, the original CNI scenario has been excluded by the CMB observations~\cite{Adshead:2013qp,Adshead:2013nka}.
However, if the ALP coupled with the gauge fields is a spectator field, the dynamics of the ALP and gauge fields is expected to be irrelevant to the scalar perturbation and then the observable chiral gravitational waves can be predicted without spoiling the success of inflation~\cite{Dimastrogiovanni:2016fuu}.
In this case, the contribution of the gauge field perturbations can dominate the primordial gravitational waves, resulting in chiral and non-Gaussian gravitational waves~\cite{Dimastrogiovanni:2016fuu,Agrawal:2017awz,Agrawal:2018mrg,Dimastrogiovanni:2018xnn,Fujita:2018vmv,Fujita:2021flu,Ishiwata:2021yne}.
This is contrary to the single-field slow-roll inflation models, where the primordial gravitational waves are originated from the quantum fluctuations of the metric itself, which are (almost) Gaussian, non-chiral, and related to the energy scale of inflation.
Therefore, the ALP-gauge-fields dynamics is also important in that it predicts nonstandard gravitational wave or $B$-mode signals.

In a previous work with two of the current authors~\cite{Fujita:2021eue}, we considered the extension of the CNI model with the SU($N$) gauge group, ``SU($N$)-natural inflation'' and provided a general procedure for constructing homogeneous isotropic attractor solutions of the gauge fields.
We found that there are multiple solutions with different background amplitudes corresponding to different spontaneous symmetry breaking patterns of the SU($N$) gauge group.
The gauge fields have nonzero background in the SU(2) subalgebra, which is broken with the spatial rotation SO(3) symmetry into the diagonal SO(3) symmetry as in the CNI model.
We also numerically simulated the dynamics of the gauge fields and showed that the analytically constructed solutions can explain all of the numerical solutions.
However, the effects of the different background solutions on the metric and gauge field perturbations were not investigated there.
Since the background solution has a similar configuration to in the CNI model, we expect that SU($N$)-natural inflation also show the enhancement of the gauge field perturbations and the generation of the gravitational waves as in the CNI model.
Notably, the original CNI model is observationally excluded, and a na\"{i}ve expectation infers that the existence of multiple vacuum configurations corresponding to the different symmetry breaking patterns might enlarge the viable parameter space compared to CNI -- we may refer to Eq.~\eqref{Ph ratio} for reasoning. In this respect, the dynamics of the perturbations in SU($N$)-natural inflation is important in that it determine the validity of this model as an inflationary model or a model predicting observable gravitational waves.

In this paper, we study the linear perturbations in SU($N$)-natural inflation.
Although the gauge fields have additional degrees of freedom in SU($N$)-natural inflation compared to the CNI model, we can classify the gauge field perturbations according to the representation of the SU(2) subalgebra.
As a result, the additional components of the gauge field perturbations are decoupled from the metric perturbations at the linear level.
Thus, the gauge field perturbations in the SU(2) subsectors that spontaneously break to the single SO($3$) are the only ones that couple with the metric perturbations, and the gravitational waves are enhanced in essentially the same way as the CNI model.
In this sense, we show a ``universality'' of the linear perturbations in SU($N$)-natural inflation.
The resultant predictions at the tree level are identical to those in CNI, which is rather contrary to our initial speculation described in the previous paragraph.
In order to break this degeneracy, we also discuss a possible transition of the background solutions due to the non-stationary behavior of the axion field and mention an expected signature in the gravitational wave power spectrum.

This paper is organized as follows.
In section~\ref{sec: model and background}, we introduce the model of SU($N$)-natural inflation and briefly summarize the background dynamics of the gauge fields.
Then, we consider the superposition of the background solutions and see that such a solution can be discussed in the context of a single solution in section~\ref{sec: multi-SU(2)}.
In section~\ref{sec: universality of linear perturbation}, we show the universality of the linear perturbations in SU($N$)-natural inflation and discuss the possible transition of the background solutions and its effects.
Section~\ref{sec: summary and discussion} is devoted to the summary and discussion of our results.
In Appendix \ref{app: gauge field perturbations}, we derive the quadratic actions of the additional subsectors and explicitly demonstrate their decoupling from the SU($2$) subgroup and from each other in the case of SU($3$) gauge fields.

%%%%%%%%%%%%%%%%%%%%%%%%%%%%%%%%%%%%%%%%%%%%
\section{Model and background solutions}
\label{sec: model and background}
%%%%%%%%%%%%%%%%%%%%%%%%%%%%%%%%%%%%%%%%%%%%

We consider a pseudo-scalar field $\phi$ coupled with SU($N$) gauge fields $A_\mu^a$ through the Chern--Simons coupling, together driving inflation:
\begin{align}
    \mathcal{L}
    =
    -\frac{1}{4}F_{\mu\nu}^a F^{a \mu\nu} 
    +\frac{1}{2}\partial_\mu \phi \partial^\mu \phi
    -V(\phi)
    +\frac{\phi}{4f}F_{\mu\nu}^a \tilde{F}^{a \mu\nu}.
\end{align}
The field strength of the SU($N$) gauge fields $F^a_{\mu\nu}$ and its dual $\tilde{F}^{a\mu\nu}$ are defined by
\begin{align}
    F^a_{\mu\nu}
    &\equiv
    \partial_\mu A^a_\nu - \partial_\nu A^a_\mu - g f^{a b c} A^b_\mu A^c_\nu,
\\
    \tilde{F}^{a \mu \nu}
    &\equiv
    \frac{\epsilon^{\mu\nu\rho\sigma}}{2} \, F^a_{\rho \sigma},
\end{align}
where $f^{a b c}$ is the structure constant of the SU($N$) algebra, the superscripts $a,b,c$ run from $1$ to $N^2-1$, $g$ is the gauge coupling constant, 
and $\epsilon^{\mu\nu\rho\sigma}$ is the totally antisymmetric tensor compatible with the spacetime metric.
The above expressions are in the component form with respect to the SU($N$) index, but equivalently the gauge fields can also be expressed by contracting with the SU($N$) generators $T^a$ as $A_\mu = A_\mu^a T_a$.
We assume that the background metric during inflation is
described by the flat FLRW metric:
\begin{equation}
\label{FLRWmetric}
    \mathrm{d}s^2 
    =
    \mathrm{d} t^2 - a(t)^2  \mathrm{d}\bm{x}^2
    =
    a(\tau)^2 \qty( \mathrm{d} \tau^2 - \mathrm{d}\bm{x}^2 )\; ,
\end{equation}
where $t$ and $\tau$ are the physical and conformal time, respectively.
Throughout this paper, we assume that the de Sitter limit is relevant for our computation during inflation, and thus $a \simeq \exp(Ht) \simeq - 1 / (H \tau)$ with constant $H$, up to appropriate integration constants.
Note that a particular case of the $N=2$ gauge group is known as the chromo-natural inflation (CNI) \cite{Adshead:2012kp}, and it has several conceptually favorable features while facing a critical observational problem \cite{Dimastrogiovanni:2012ew, Adshead:2013nka}.

In this section, we consider the background solution of the gauge fields $\bar{A}_\mu^a$ and inflaton $\bar{\phi}$, 
both of which depend only on time to be compatible with the background spacetime \eqref{FLRWmetric}.
In the previous study~\cite{Fujita:2021eue},
it has been shown through the analytical discussion and numerical simulations that the gauge fields have background solutions corresponding to SU(2) subalgebras in SU($N$), dynamically realizing a configuration compatible with the background metric \eqref{FLRWmetric}.
The non-zero background of the gauge fields spontaneously breaks the spatial rotation symmetry and the global gauge symmetry of the SU(2) subgroup into the diagonal SO(3), namely SO(3) $\times$ SU(2) $\to$ SO(3).

\begin{figure}[tbp]
  \begin{center}
  \includegraphics[width=0.6\textwidth]{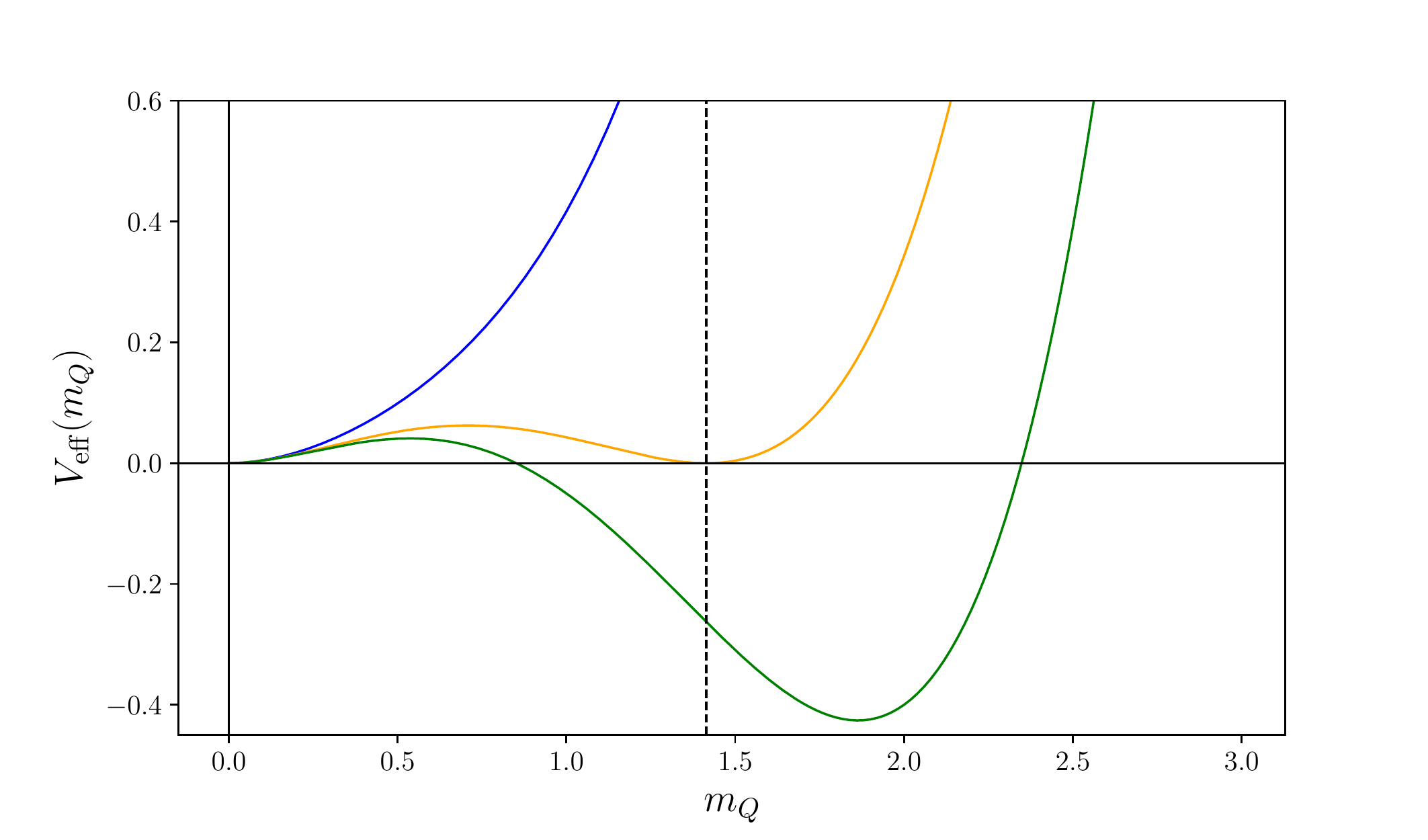}
  \end{center}
  \caption
 {Effective potential for the gauge field vev $m_Q$ defined in Eq.~\eqref{Veff_def} for $\xi=1$ (blue), $3/\sqrt{2}$ (orange), and $2.4$ (green). One can observe that a non-trivial stable solution at $m_Q=m_{Q+}$ becomes the true vacuum when $\xi>3/\sqrt{2}$. In the marginal case with $\xi=3/\sqrt{2}$, we have a degenerated vacuum at $m_{Q+}=\sqrt{2}$ (vertical dashed line). Thus, the non-trivial true vacuum always appears at $m_{Q+}>\sqrt{2}$.
 Note that $V_{\rm eff}$ is an effective potential only for the dynamics of $Q$ and the apparent negative value in the green curve has no physical consequence for the expansion.}
 \label{fig:Veff}
\end{figure}

In this paper, we focus on this class of configurations of the gauge fields
for each SU($2$) subgroup of a parent SU($N$) and their superpositions and assume that these configurations are realized as stable solutions with an appropriate choice of the parameters.
First, we consider the configurations corresponding to a single SU(2) subalgebra and discuss the superpositions later.
Here, we adopt the temporal gauge $\bar{A}_0^a(t) = 0$,%
\footnote{In fact, this gauge condition can as well result from the constraint equation obtained by varying the background action with respect to $\bar{A}_0$ and does not completely fix the gauge freedom. 
Nevertheless, for our purpose this condition together with the ansatz \eqref{eq: BG Ai formula} sufficiently removes the gauge ambiguity.}
and then the gauge field configuration of our interest can be written as 
\begin{align}
    \bar{A}_0(t) 
    =
    0,
    \quad
    \bar{A}_i(t) 
    \equiv
    a(t) Q(t) \mathcal{T}_i,
    \label{eq: BG Ai formula}
\end{align}
where $\mathcal{T}_i$ represents the generator of the SU($2$) subalgebra.
We construct $\mathcal{T}_i$ from the total SU($N$) generator $T_a$ such that they satisfy $\Tr \qty[\mathcal{T}_i \mathcal{T}_j] = \delta_{i j}/2$ and $\qty[ \mathcal{T}_i,\mathcal{T}_j ] = i \lambda \epsilon^{i j k} \mathcal{T}_k$.
Note that we can always rearrange the SU(2) generators so that $\bar{A}_i \propto \mathcal{T}_i $ for the isotropic configuration.
Since this definition leaves the normalization of $\mathcal{T}_i$ the same as the original SU($N$) generator $T_a$,
a coefficient $\lambda$ appears in the SU(2)-like commutation relation inside the corresponding subalgebra,
depending on which SU($2$) subgroup breaks down by acquiring a vev.
If the parent gauge group is already SU($2$),
SU(2) gauge fields only take a nontrivial configuration \eqref{eq: BG Ai formula} with $\lambda=1$.
However,  SU(N$>2$) gauge fields can take configurations with $\lambda$ which is generally different from
unity and is determined by the choice of the SU($2$) subalgebra.
For example, SU(3) gauge fields can take either of two stable configurations of the SU(2) subalgebras with $\lambda=1$ or $1/2$. In this sense, one can regard $\lambda$ as the label of different gauge field solutions, corresponding to different symmetry breaking patterns, at the background level.
Using the ansatz~\eqref{eq: BG Ai formula}, the equations of motion (EoMs) for $\bar{\phi}$ and $\bar{A}_i$ become
\begin{align}
    \ddot{\bar{\phi}} + 3 H \dot{\bar{\phi}} + \partial_\phi V(\bar{\phi})
    &=
    -\frac{3 g \lambda}{f} Q^2 \qty( \dot{Q} + H Q ),
    \label{eom_phi}
    \\
    \ddot{Q} + 3 H \dot{Q} + \left( 2 H^2 + \dot{H} \right) Q + 2 g^2 \lambda^2 Q^3
    &=
    \frac{g \lambda}{f} Q^2 \dot{\bar{\phi}},
    \label{eom_Q}
\end{align}
where the dots represent derivatives with respect to the physical time $t$.
In the following, we 
seek for a solution with which the gauge fields have nonzero background and balance with 
$\bar\phi$'s velocity to realize slow-roll inflation without necessitating a flat potential $V$.
Moreover, as the leading order in slow roll, we take the de Sitter limit $\dot{H} = 0$, focus on static solutions with $\dot{Q}=0$, and introduce a dimensionless gauge field amplitude $m_Q$ defined by
\begin{align}
    m_Q \equiv \frac{g \lambda }{H}Q.
\end{align}
Under these conditions, we can parametrize $\bar{\phi}$'s velocity by a dimensionless parameter $\xi \equiv \dot{\bar{\phi}}/(2 f H) \simeq \mathrm{const}.$ Then the EoM for $Q$ leads to
\begin{equation}
    m_Q^3 - \xi m_Q^2 + m_Q = 0.
    \label{eq: EoM for mQ}
\end{equation}
The solutions are 
\begin{equation}
    m_Q 
    =
    0,  m_{Q +}, m_{Q -},
    \label{eq: mQ determined by xi}
\end{equation}
with
\begin{equation}
    m_{Q \pm}\equiv \frac{\xi \pm \sqrt{\xi^2-4}}{2},
\end{equation}
when $\xi > 2$.
From the EoM~\eqref{eq: EoM for mQ}, we introduce the effective potential for $m_Q$ as
\begin{equation}
    V_\mathrm{eff}(m_Q)
    =
    \frac{1}{2}m_Q^2 
    - \frac{1}{3}\xi m_Q^3 
    + \frac{1}{4} m_Q^4,
    \label{Veff_def}
\end{equation}
which has extrema corresponding to the solutions~\eqref{eq: mQ determined by xi}  for a fixed value of $\xi$.
Clearly $V_{\rm eff}$ in \eqref{Veff_def} is invariant under a simultaneous inversion $\xi \to - \xi$ and $m_Q \to - m_Q$, or equivalently $\dot{\bar{\phi}} \to - \dot{\bar\phi}$ and $Q \to - Q$. This potential is illustrated in Fig.~\ref{fig:Veff}.
%-----------------------FIGURE-------------------------%
%-----------------------FIGURE-------------------------%
When $\xi > 2$, $m_Q = 0$ and $m_{Q+}$ are local minima, and $m_Q = m_{Q-}$ is a local maximum.
Moreover, $V_\mathrm{eff}(m_{Q_+}) < V_\mathrm{eff}(0)$ if and only if $\xi > 3/\sqrt{2}$.
Therefore, the solution $m_Q = m_{Q+}$ becomes the unique true vacuum when $\xi > 3/\sqrt{2}$ or equivalently $m_{Q+} > \sqrt{2}$, and otherwise the trivial solution $m_Q = 0$ is the true vacuum, 
which is merely a single-scalar inflation paradigm and is out of our current interest.
Thus, in the following, we assume $\xi > 3/\sqrt{2}$ and focus on the nontrivial solution $m_Q = m_{Q+}$.
In fact, $m_{Q+}$ can also be determined through the EoM for $\bar{\phi}$ as
\begin{equation}
    m_{Q+} \simeq 
    \qty(
        \frac{- g^2 \lambda^2 f \partial_\phi V\qty(\bar{\phi})}{3 H^4}
    )^{\frac{1}{3}}.
    \label{eq: mQ detemined by potential}
\end{equation}
The fact that this solution is a dynamical attractor ensures that the potential term in \eqref{eom_phi} balances the term on the right-hand side and is the very reason why the axion potential $V(\phi)$ does not need to be flat to realize inflation.
Starting from \eqref{eq: mQ detemined by potential}, we can view the EoM~\eqref{eq: EoM for mQ} as an equation which determines $\xi$ from $m_{Q+}$ as
\begin{equation}
    \xi = m_{Q+} + \frac{1}{m_{Q+}}.
    \label{eq: xi determined by mQ}
\end{equation}
Therefore, once we fix the potential and parameters, we can confirm that the above conditions are satisfied by calculating $m_{Q+}$ and $\xi$ from Eqs.~\eqref{eq: mQ detemined by potential} and \eqref{eq: xi determined by mQ}.

Note that, in the above equations, $\lambda$ always appears in the form of $g \lambda$.
This is because the non-linear part of the field strength $F^a_{\mu\nu}\supset g[A_\mu, A_\nu]$ reads $g[\bar{A}_i, \bar{A}_j]\propto g[ \mathcal{T}_i,\mathcal{T}_j ] \propto g\lambda \, \epsilon^{ijk} \mathcal{T}_k$ at the background.
Therefore, if only one nontrivial solution realizes during the inflationary epoch (i.e.~fixed $\lambda$), the difference in the choice of the SU(2) subalgebra characterized by $\lambda$ is degenerate with that of $g$.

%%%%%%%%%%%%%%%%%%%%%%%%%%%%%%%%%%%%%%%%%%%%
\section{Multiple SU(2) subalgebras}
\label{sec: multi-SU(2)}
%%%%%%%%%%%%%%%%%%%%%%%%%%%%%%%%%%%%%%%%%%%%
So far, we have focused on the gauge field configuration corresponding to a single SU(2) subalgebra.
In this section, we consider multiple SU(2) subalgebras embedded in SU($N$) with $N \geq 4$, which can be decomposed into at least two SU($2$) subalgebras, and discuss the case where the gauge fields have nonzero background in each SU(2) subalgebra.
In this case, the gauge field configuration can be represented by 
\begin{align}
    \bar{A}_0(t) = 0,
    \quad
    \bar{A}_i(t) 
    \equiv
    a(t) \sum_{\alpha = 1}^{n} Q_{(\alpha)}(t) \mathcal{T}^{(\alpha)}_i,
    \label{eq: BG Ai formula multi-SU(2)}
\end{align}
where $\alpha$ is an index of the total $n$ SU(2) subalgebras, and $\mathcal{T}^{(\alpha)}_i$
is constructed to represent the generator of the $\alpha$-th SU($2$) subalgebra and satisfies $\mathrm{Tr} [\mathcal{T}^{(\alpha)}_i \mathcal{T}^{(\beta)}_j] = \delta_{i j}\delta^{\alpha \beta}/2$ and $[ \mathcal{T}^{(\alpha)}_i,\mathcal{T}^{(\beta)}_j ] = i \lambda_{(\alpha)} \epsilon^{i j k} \delta^{\alpha \beta} \mathcal{T}^{(\alpha)}_k$.
We again choose $\mathcal{T}_i^{(\alpha)}$ so that $\mathcal{T}_i^{(\alpha)}$ corresponds to $\bar{A}_i$ for each SU(2) subalgebra.
Apart from possibly different values of $\lambda_{(\alpha)}$,
the situation is the same as the model in which $n$ copies of the SU(2) gauge fields are coupled to an axion through the Chern--Simons term with the same $g$ and $f$. Using this configuration, the EoMs become
\begin{align}
    \ddot{\bar{\phi}} + 3 H \dot{\bar{\phi}} + \partial_\phi V(\bar{\phi})
    &=
    -\frac{3 g}{f} \sum_{\alpha = 1}^{n} \lambda_{(\alpha)} Q_{(\alpha)}^2 \qty( \dot{Q}_{(\alpha)} + H Q_{(\alpha)} ),
\\
    \ddot{Q}_{(\alpha)} + 3 H \dot{Q}_{(\alpha)} + 
    \left( 2 H^2 + \dot{H} \right) Q_{(\alpha)} + 2 g^2 \lambda_{(\alpha)}^2 Q_{(\alpha)}^3
    &=
    \frac{g \lambda_{(\alpha)}}{f} Q_{(\alpha)}^2 \dot{\bar{\phi}}.
\end{align}
In the de Sitter and slow-roll limit with $\xi \equiv \dot{\bar{\phi}}/(2 f H) = \mathrm{const}.$, we obtain
\begin{align}
    \partial_\phi V(\bar{\phi})
    &=
    -\frac{3 g H}{f} \sum_{\alpha = 1}^{n} \lambda_{(\alpha)} Q_{(\alpha)}^3,
    \label{eq: EoM for Potential gradient multi-SU(2)}
\\
    m_{Q(\alpha)} 
    &\equiv
    \frac{g \lambda_{(\alpha)} }{H}Q_{(\alpha)}
    =
    0, \; \frac{\xi \pm \sqrt{\xi^2-4}}{2}.
\end{align}
Here, we assume $\xi > 3/\sqrt{2}$ and that the gauge fields in each SU(2) subalgebra acquire a non-zero amplitude, which is generically different from one another, $Q_{(\alpha)}=(\xi+\sqrt{\xi^2-4})H/(2g\lambda_{(\alpha)})$. 
However, the amplitudes look identical in $m_{Q(\alpha)} = m_{Q+} \equiv (\xi + \sqrt{\xi^2-4})/2 $.
Then, the sum of the background gauge fields in \eqref{eq: BG Ai formula multi-SU(2)} become
\begin{align}
    \bar{A}_i(t) 
    =
    \frac{a H m_{Q+}}{g} \sum_{\alpha = 1}^{n} \frac{\mathcal{T}^{(\alpha)}_i}{\lambda_{(\alpha)}}
    \equiv 
    \frac{a H m_{Q+}}{g \hat{\lambda}} \hat{\mathcal{T}}_i,
\end{align}
where
\begin{align}
\label{def_hatT}
    \hat{\mathcal{T}}_i
    \equiv
    \hat{\lambda} \sum_{\alpha = 1}^{n} \frac{\mathcal{T}^{(\alpha)}_i}{\lambda_{(\alpha)}},
    \quad
    \hat{\lambda}
    \equiv 
    \qty(\sum_{\alpha = 1}^{n} \frac{1}{\lambda_{(\alpha)}^2})^{-1/2},
\end{align}
so that they satisfy $\mathrm{Tr} [\hat{\mathcal{T}}_i \hat{\mathcal{T}}_j] = \delta_{i j}/2$ and $[ \hat{\mathcal{T}}_i , \hat{\mathcal{T}}_j ] = i \hat{\lambda} \epsilon^{i j k} \hat{\mathcal{T}}_k$.
In other words, $\hat{\mathcal{T}}_i/\hat{\lambda}$ is a generator of an SU(2) subalgebra,
whereas $\hat{\mathcal{T}}_i$ is a linear combination of all the SU($2$) subalgebras.
Then Eq.~\eqref{eq: EoM for Potential gradient multi-SU(2)} leads to
\begin{equation}
    m_{Q+} 
    = 
    \qty(
        \frac{- g^2 \hat{\lambda}^2 f \partial_\phi V\qty(\bar{\phi})}{3 H^4}
    )^{\frac{1}{3}}.
\end{equation}
These results are the same as the case of a single SU(2) subalgebra
in Sec.~\ref{sec: model and background} with the replacement $\{\mathcal{T}_i,\lambda\}\to \{\hat{\mathcal{T}}_i, \hat{\lambda}\}$. 
This simple result is a consequence of the fact that each of $n$ SU($2$) subalgebras, together with a single SO($3$) of the background spacetime, is broken as SO($3$) $\times$ SU($2$) $\to$ SO($3$). 
Therefore, we can relate the background solution of the gauge fields to a single SU(2) subalgebra.
Thanks to this fact, it suffices to consider the case of a single SU($2$) subalgebra with $\mathcal{T}_i$ and $\lambda$, and the results apply to any multiple SU($2$) case simply by the replacement of $\mathcal{T}_i \to \hat{\mathcal{T}}_i$ and $\lambda \to \hat\lambda$. In the following, we show that this property propagates to the analysis of linear perturbations.

%%%%%%%%%%%%%%%%%%%%%%%%%%%%%%%%%%%%%%%%%%%%
\section{Universality of linear perturbation}
\label{sec: universality of linear perturbation}
%%%%%%%%%%%%%%%%%%%%%%%%%%%%%%%%%%%%%%%%%%%%

In this section, we consider the quadratic action of the perturbations and show the universality of
the observable predictions for linear perturbations in SU($N$)-natural inflation:
the predictions for the linear metric perturbations in SU($N$)-natural inflation are indistinguishable irrespective of the gauge symmetry SU($N$) or the background configuration of the gauge fields characterized by $\lambda$, as long as $\lambda$ is constant during inflation.
The essential reasons for this result can be summarized as follows:
\begin{itemize}
    \item The $\bm{3}$ representation of the SU(2) subalgebra corresponding to the background solution behaves as the $\bm{3}$ representation of the diagonal SO(3).
    \item Due to this correspondence, only the gauge field perturbations in the $\bm{3}$ representation can linearly couple with the metric perturbations and the gauge field perturbations in each SU(2) representation decouples from each other at the linear level.
    \item The information about the background configuration $\lambda$ arises only through the nonlinear interactions of non-Abelian gauge fields and at the linear level enters as, schematically, ${\rm Tr}[F^2] \sim g {\rm Tr}[A^2 \partial A], \, g^2 {\rm Tr}[A^4] \sim g \lambda Q \delta A \partial \delta A, \, g^2 \lambda^2 Q^2 \delta A^2$, and therefore $\lambda$ always appears in the combination $g \lambda$.
\end{itemize}
To show this fact more concretely, we decompose the gauge fields according to the SU(2) representations as
\begin{align}
    A_i 
    =
    \left(\bar{A}_i^I + \delta A_i^I\right) \mathcal{T}_I
    + \delta A_i^A \tilde{\mathcal{T}}_A,
    \label{eq: gauge field decomposition}
\end{align}
where $I$ runs from $1$ to $3$, and $A$ runs from $4$ to $N^2-1$.
Here, $\mathcal{T}_I$ is the same as the one in Eq.~\eqref{eq: BG Ai formula} (or $\hat{\mathcal{T}}_i$ in \eqref{def_hatT} in the multiple SU(2) case).
$\tilde{\mathcal{T}}_A$ represents the other SU($N$) generators, which are further decomposed according to their SU(2) representation.
Although $I$ is an index of the generators and different from a spatial index $i$, we can identify $I$ as a spatial index in the presence of the background gauge fields for the aforementioned reason. 
We thus do not discriminate between these indices $i$ and $I$ hereafter. Then part of the structure constants can be replaced by
\begin{align}
\label{structure_property}
    f^{i j k} = \lambda \epsilon^{i j k},
    \quad
    f^{i j A} = 0,
\end{align}
since $\mathcal{T}_i$ generates the SU(2) subalgebra.

First, we consider the linear coupling between the gauge field and metric perturbations.
The gauge field perturbations can couple with the metric perturbations through the kinetic term of the gauge fields, $- g^{\mu \nu} g^{\rho \sigma} F_{\mu \rho}^a F_{\nu \sigma}^a/4$.
In order to determine possible linear couplings, we consider the linear terms of the gauge field perturbations in $F_{\mu \rho}^a F_{\nu \sigma}^a$.
Since the background field strength $\bar{F}_{\mu \rho}^a$ has a non-zero value only in the $\bm{3}$ component $\bar{F}_{\mu \rho}^i$ (i.e. $\bar{F}^A_{\mu\rho}=0$), 
we can decompose $F_{\mu \rho}^a F_{\nu \sigma}^a$ as
\begin{align}
    F_{\mu \rho}^a F_{\nu \sigma}^a
    =
    \delta F_{\mu \rho}^i \bar{F}_{\nu \sigma}^i
    + \bar{F}_{\mu \rho}^i \delta F_{\nu \sigma}^i
    + (\mathrm{higher \,\, order \,\, terms}),
\end{align}
where $\delta F_{\mu \rho}^i$ is the linear perturbation in $F_{\mu \rho}^i$.
Moreover, $\delta F_{\mu \rho}^i$ is given by
\begin{align}
    \delta F_{\mu \rho}^i
    =
    \partial_\mu \delta A_\rho^i
    - \partial_\rho \delta A_\mu^i
    - g \lambda \epsilon^{i j k} 
    \left( 
        \bar{A}_\mu^j \delta A_\rho^k
        +
        \delta A_\mu^j \bar{A}_\rho^k
    \right),
\end{align}
which includes only the gauge field perturbations in the $\bm{3}$ representation, $\delta A_\mu^i$,
thanks to the properties \eqref{structure_property}.
Therefore, the metric perturbations can linearly couple only with the gauge field perturbations in the $\bm{3}$ representation.

Next, we discuss the couplings between the gauge field perturbations.
They can couple with each other through the kinetic term and Chern--Simons coupling term.
Then, we consider $F_{\mu \rho}^a F_{\nu \sigma}^a$ again, since its different contractions yield these terms.
The field strength is decomposed as
\begin{align}
    F_{\mu \rho}^i
    &=
    \partial_\mu (\bar{A}_\rho^i + \delta A_\rho^i)
    - \partial_\rho ( \bar{A}_\mu^i + \delta A_\mu^i)
    - g \lambda \epsilon^{i j k} 
    (\bar{A}_\mu^j + \delta A_\mu^j)
    (\bar{A}_\rho^k + \delta A_\rho^k)
    - g f^{i B C} \delta A_\mu^B \delta A_\rho^C
\nonumber\\
    &=
    \bar{F}_{\mu \rho}^i 
    + \mathcal{O}(\delta A_\mu^j)
    + \mathcal{O}(\delta A_\mu^j \delta A_\rho^k)
    + \mathcal{O}(\delta A_\mu^B \delta A_\rho^C),
    \\
    F_{\mu \rho}^A
    &=
    \partial_\mu \delta A_\rho^A
    - \partial_\rho \delta A_\mu^A
    - g f^{A j B} (\bar{A}_\mu^j + \delta A_\mu^j) \delta A_\rho^B
    - g f^{A B j} \delta A_\mu^B (\bar{A}_\rho^j + \delta A_\rho^j)
    - g f^{A B C} \delta A_\mu^B \delta A_\rho^C
\nonumber\\
    &=
    \partial_\mu \delta A_\rho^A
    - g f^{A j B} \bar{A}_\mu^j \delta A_\rho^B
    - ( \mu \leftrightarrow \rho)
    + \mathcal{O}(\delta A^2).
\end{align}
As a result, we obtain
\begin{align}
    F_{\mu \rho}^a F_{\nu \sigma}^a
    =
    \bar{F}_{\mu \rho}^i \bar{F}_{\nu \sigma}^i
    + \mathcal{O}(\delta A_\mu^j)
    + \mathcal{O}(\delta A_\mu^j \delta A_\rho^k)
    + \mathcal{O}(\delta A_\mu^B \delta A_\rho^C) \; ,
\end{align}
up to the quadratic orders.
Therefore, the perturbations in the $\bm{3}$ representation are decoupled from the other representations.
Moreover, we also find that the other representations, if there are multiple of them, are decoupled from each other for the following reason.
$F_{\mu \rho}^a F_{\nu \sigma}^a$ has two types of the $\mathcal{O}(\delta A_\mu^B \delta A_\rho^C)$ terms; one is proportional to $\delta A_\rho^A \delta A_\sigma^A$ and the other is proportional to $f^{i B C} \delta A_\mu^B \delta A_\rho^C$.
It is obvious that the former couples within the same representation.
This is also true for the latter since $f^{i B C}$ describes the action of the SU(2) subalgebra on the representation labelled by $B$ and takes a nonzero value only when $B$ and $C$ are the labels of the same representation (see Eqs.~\eqref{TE_commutator} and \eqref{TT_commutator} for some concrete examples).

From the above discussion, we have shown that only the gauge field perturbations in the $\bm{3}$ representation can linearly couple with the metric perturbations and that the gauge field perturbations in each representation decouples from each other at the linear level.
Therefore, it is sufficient to focus on the $\bm{3}$ representation to investigate the effect of the gauge field perturbations on the metric perturbations, which in the end connect to the observables.

As long as we focus on the $\bm{3}$ representation, the relevant degrees of freedom of the gauge field is the same as the CNI model, and then the dynamics of the gauge fields and metric is 
practically the same as the CNI model.
The only difference is the replacement of $g$ with $g \lambda$ corresponding to the choice of the background solutions.
Thus, with this replacement, the observable predictions of SU($N$)-natural inflation are identical to those of the CNI model.
Unless one {\it a priori} knows the value of the gauge coupling constant $g$, $g \lambda$ should be taken as one parameter and we cannot distinguish these models of inflation.
In other words, there is a universality of the linear perturbations in SU($N$)-natural inflation.
Though beyond the scope of our present study, we would need to go into higher orders to see observable differences, since the gauge field perturbations in the representations other than the $\bm{3}$ representation can affect the metric perturbations at the nonlinear level or through loops.
The example of the quadratic action of the gauge field perturbations in the SU(3) case is studied in App.~\ref{app: gauge field perturbations}.

%%%%%%%%%%%%%%%%%%%%%%%%%%%%%%%%%%%%%%%%%%%%
\subsection{Transition of the solutions}
\label{subsec: transition}
%%%%%%%%%%%%%%%%%%%%%%%%%%%%%%%%%%%%%%%%%%%%

We have shown the universality in SU($N$)-natural inflation.
Yet, as stated, it assumes that the background solution is unique throughout inflation,
i.e.,~constant $\lambda$. 
However, this parameter merely labels which of the symmetry breaking patterns of SU($N$) is dynamically realized under the assumptions of stationary solutions, and in reality it faces the dynamics of the whole system.
If the gauge field background experiences the transition during inflation, the degeneracy between $g$ and $\lambda$ can be broken due to the change of the $\lambda$ value, and then we cannot apply the analysis of the CNI model.
Here, we consider the possibility of transitions of the gauge field configuration in SU($N$)-natural inflation, which may be an important caveat to the universality of the CNI prediction.

First, we discuss how such a transition can occur.
In the previous section, we assumed $\xi \equiv \dot{\bar{\phi}}/(2 f H) = \mathrm{const}.$
However, in a realistic situation, $\xi$ and $m_{Q+}$ evolve in time.
Here, we choose the cosine-type potential as a typical example of the ALP potential:
\begin{align}
    V(\phi) = \mu^4 \qty[ 1 + \cos \qty(\frac{\phi}{F}) ],
    \label{eq: cosine potential}
\end{align}
where $\mu$ is the scale of the potential, and $F$ is a decay constant of the ALP.
The following discussion can also be applied to other potentials.
As discussed in Sec.~\ref{sec: model and background}, whether each background solution can be realized or not is determined by the condition $m_{Q+} \gtrless \sqrt{2}$.
Then, we evaluate $m_{Q+}$ through Eq.~\eqref{eq: mQ detemined by potential} with a fixed $\lambda$ and check the condition $m_{Q+} > \sqrt{2}$.
By using the potential~\eqref{eq: cosine potential} and the approximated Friedmann equation $3 M_\mathrm{Pl}^2 H^2 \simeq V(\bar \phi)$, 
we obtain the dependence of $m_{Q+}$ on $\bar{\phi}$ and $\lambda$ as
\begin{align}
    m_{Q+}^3 (\lambda)
    \propto
    - \lambda^2 \frac{\partial_\phi V(\bar{\phi})}{V^2(\bar{\phi})}
    \propto
    \lambda^2 \frac{\sin \qty(\frac{\bar{\phi}}{F})}{\qty[1 + \cos\qty(\frac{\bar{\phi}}{F})]^2},
\end{align}
which is a monotonically increasing function in $0 < \bar{\phi}/F < \pi$.
Therefore, the solution with smaller $\lambda$ becomes a possible solution as 
$\bar\phi$ rolls down the potential.
Especially, if $m_{Q+} \geq \sqrt{2}$ is not initially satisfied even for $\lambda = 1$, the isotropic gauge field background can emerge in the course of inflation as in the SU(2) case~\cite{Domcke:2018rvv}.

On the other hand, if
the ALP is a spectator field, the Hubble parameter may be taken approximately constant, and then the dependence of $m_{Q+}$ on $\bar{\phi}$ and $\lambda$ is given by
\begin{align}
    m_{Q+}^3 (\lambda)
    \propto
    - \lambda^2 \partial_\phi V(\bar{\phi})
    \propto
    \lambda^2 \sin \qty(\frac{\bar{\phi}}{F}),
\end{align}
which is a monotonically increasing function in $0 < \bar{\phi}/F < \pi/2$ and a monotonically decreasing function in $\pi/2 < \bar{\phi}/F < \pi$.
Therefore, the range of possible $\lambda$ values is once broadened and then becomes narrower as the spectator ALP rolls down the potential from $0 < \bar{\phi}/F < \pi/2$.

From the above, we can see that the range of possible values of $\lambda$
that realizes stable configurations may change as the ALP rolls down the potential, which 
potentially induces the transition of the gauge field configuration.
Although a hint of such a transition has been shown previously \cite{Fujita:2021eue}, there is room to understand how it occurs as an entire system. This question should be addressed by a numerical simulation of the dynamical system of the background metric, ALP, and gauge fields in future publications, but this is out of the scope of this paper.

Before closing the section, we consider how the change of the $\lambda$ value could affect the linear perturbations.
The tensor component of the gauge field perturbations experiences a tachyonic instability in the presence of the gauge field background, and they source the tensor component of the metric perturbations or the gravitational wave, whose amplitude exponentially depends on the $\lambda$ value.
Thus, we expect that the tensor perturbations most significantly depend on the $\lambda$ value for a fixed value of $g$.
The power spectrum of the sourced gravitational waves is given by 
\begin{align}
    \frac{k^3}{2 \pi^2} P_h^{\mathrm{sourced}}
    =
    \frac{\epsilon_B H^2}{\pi^2 M_\mathrm{Pl}^2} |\mathcal{F}(m_{Q+})|^2,
    \label{Phsourced}
\end{align}
where $\epsilon_B \equiv g^2 \lambda^2 Q^4/(M_\mathrm{Pl}^2 H^2)$ and $|\mathcal{F}(m_{Q+})| \simeq e^{2.4 m_{Q+}}$, whose exact expression is given in~\cite{Dimastrogiovanni:2016fuu}.
To evaluate the effect of the transition on the gravitational waves, let us compare their power spectrum by fixing the parameters other than $\lambda$ based on Eq.~\eqref{Phsourced}.
The $\lambda$-dependence of $P_h^{\mathrm{sourced}}$ is given by
\begin{equation}
    \frac{P_h^{\mathrm{sourced}}(\lambda)}{P_h^{\mathrm{sourced}}(\lambda=1)}
    \simeq \lambda^{2/3} \exp[4.8(\lambda^{2/3}-1)\bar{m}_{Q+}],
    \label{Ph ratio}
\end{equation}
where $\bar{m}_{Q+}\equiv m_{Q+}/\lambda^{2/3}$ is independent of $\lambda$ and we assume that the gauge fields make a subdominant contribution to the total energy density.
Thus, as $\lambda$ decreases, the amplitude of the sourced gravitational waves rapidly drops. 
For a larger $N$ of SU($N$), the gauge fields can take a configuration with a smaller $\lambda=\sqrt{6/(N(N^2-1))}$~\cite{Fujita:2021eue}.
For instance, SU($N\ge 4$) can take $\lambda=1/\sqrt{10}$, and then the ratio
in Eq.~\eqref{Ph ratio} reads $2\times 10^{-4}$ for $\bar{m}_{Q+}=3$.
This would indeed break the degeneracy of the prediction between CNI and the SU($N>2$)-natural inflation,
since the tensor sector is exponentially sensitive to the change in $\lambda$ while the scalar sector does not experience such an exponential grow due to the gauge field production.
Therefore, if such a step-like behavior should be observed in the gravitational wave power spectrum, it would be a distinctive signal of SU($N$)-natural inflation.

%%%%%%%%%%%%%%%%%%%%%%%%%%%%%%%%%%%%%%%%%%%%
\section{Summary and discussion}
\label{sec: summary and discussion}
%%%%%%%%%%%%%%%%%%%%%%%%%%%%%%%%%%%%%%%%%%%%

In this paper, we have discussed the linear perturbations in the SU($N$)-natural inflation.
There are multiple gauge field background solutions corresponding to a choice of an SU(2) subalgebra in SU($N$), which can be labeled by a constant $\lambda$.
Thanks to the simultaneous breaking of the symmetry SO(3) $\times$ SU(2) $\to$ SO(3), the gauge field perturbations corresponding to the SU(2) generators can be regarded as a spatial tensor and they linearly couple with the metric perturbations as in the CNI model.
On the other hand, the other components of the gauge field perturbations do not couple with the metric perturbations due to the absence of the corresponding background gauge field and they decouple from each other according to the SU(2) representations.
As a result, the predictions for the metric perturbations become the same as in the CNI model except for the replacement of $g$ by $g \lambda$ if a unique background solution is realized throughout inflation.
In this case, the difference of the background solution or $\lambda$ can be compensated by the change of the coupling constant $g$.
In the observational perspective, $g$ and $\lambda$ are degenerate, and all the SU($N$)-natural models predict the same $n_s$ vs $r$ values at the tree level. In this sense, we have shown a ``universality'' of predictions for linear perturbations in SU($N$)-natural inflation.

This conclusion holds even if multiple SU($2$) subalgebras in the parent SU($N$) contributes to the background. For, one linear combination of their generators behaves as a single effective SU($2$) with a new but unique $\hat\lambda$, and the decomposition of the perturbations is necessarily done in accordance with this effective representation. The same result as above then follows for the effective SU($2$) and $\hat\lambda$.

We however caution that this does not necessarily imply a complete indistinguishability of SU($N$)-natural inflation from CNI.
If there were transitions of the background solution during inflation, the metric perturbations, especially the gravitational waves, would show features that represent a change of $\lambda$.
Such behaviors depict transitions of vacua during inflation that are controlled by the homogeneous dynamics of a considered specific SU($N$) model, which can therefore be a distinctive signal of SU($N$)-natural inflation.
Whether such transitions can occur in a realistic situation will be addressed by numerical simulations, which is left for future work.

Moreover, the loop contributions of the gauge field perturbations in the representations other than the $\bm{3}$ representation can break the universality.
The loop contributions can change the prediction in the $n_s$--$r$ plane and induce the non-Gaussianity of the metric perturbations.

Another possible direction for future research is an extension of the SU($N$)-natural inflation model.
For example, we can consider a model with multiple gauge groups such as SU($N$) $\times$ SU($M$).
In this case, each of SU($N$) and SU($M$) gauge fields can have a nonzero background in an SU(2) subalgebra and both of them affect the dynamics of the same ALP field (or multiple ALPs, in which case the universality would be trivially broken).
Since the SU($N$) and SU($M$) gauge groups have different gauge coupling in general, the unification of multiple SU(2) subalgebras into a single SU(2) subalgebra discussed in Sec.~\ref{sec: multi-SU(2)} cannot be straightforwardly applied to this case.
Thus, this extension could lead to some interesting modification of the universality.
We can also consider the effect of matter components coupled with the SU($N$) gauge fields.

%%%%%%%%%%%%%%%%%%%%%%%%%%%%%%%%%%%%%%%%%%%%
\section*{Acknowledgements}
%%%%%%%%%%%%%%%%%%%%%%%%%%%%%%%%%%%%%%%%%%%%
We would like to thank Antonio De Felice, Eiichiro Komatsu, and Misao Sasaki for useful discussions and comments.
This work is supported by the Grant-in-Aid for Scientific Research Fund of the JSPS 18K13537 (T.F.) and 20J20248 (K.M.).
K.M. is supported by World Premier International Research Center Initiative (WPI Initiative), MEXT, Japan, and the Program of Excellence in Photon Science.

\appendix

%%%%%%%%%%%%%%%%%%%%%%%%%%%%%%%%%%%%%%%%%%%%
\section{Rest of the perturbations of the gauge fields}
\label{app: gauge field perturbations}
%%%%%%%%%%%%%%%%%%%%%%%%%%%%%%%%%%%%%%%%%%%%

In Sec.~\ref{sec: universality of linear perturbation}, we have shown the decoupling of the linear perturbations according to the representations of the SU(2) subalgebra.
Here, we consider $N = 3$ as a concrete example and
explicitly show the decoupling of the gauge field perturbations.
In SU(3)-natural inflation, the gauge fields have two background solutions corresponding to the symmetry breaking pattern shown in Tab.~\ref{tab: decomposition}.

%%%%%%%%%%%%%%%%%%%%%%%%%%%%%%%%%%%%%%%
\begin{table*}[htbp]
    \centering
    \caption{Decomposition of the fundamental and adjoint representations of SU(3)~\cite{ramond_2010}.
    } 
    \label{tab: decomposition}
    \begin{tabular}{c c c c}
        case & subgroup & $\mathbf{N}$ & $\mathbf{N^2-1}$ 
        \\
        \hline
        (a) & SU(2)$\times$U(1) & $\bm{2}_{-1} + \bm{1}_2$ & $\bm{3}_0 + \bm{2}_3 + \bm{2}_{-3} + \bm{1}_0$
        \\
        %\cline{2-4}
        (b) & SU(2)& $\bm{3}$ & $\bm{3} + \bm{5}$
        \\
    \end{tabular}
\end{table*}
%%%%%%%%%%%%%%%%%%%%%%%%%%%%%%%%%%%%%%%

%%%%%%%%%%%%%%%%%%%%%%%%%%%%%%%%%%%%%%%%%%%%
\subsection{Case (a)}
\label{subsec: case a}
%%%%%%%%%%%%%%%%%%%%%%%%%%%%%%%%%%%%%%%%%%%%

First, we discuss the perturbations around the background solution in case (a).

%%%%%%%%%%%%%%%%%%%%%%%%%%%%%%%%%%%%%%%%%%%%
\subsubsection{Decomposition of the adjoint representation}
\label{subsubsec: decomposition in (a)}
%%%%%%%%%%%%%%%%%%%%%%%%%%%%%%%%%%%%%%%%%%%%

In this case, we consider the embeddings of SU($3$)$\supset$SU($2$)$\times$U($1$) and then the adjoint representation of SU($3$) is decomposed as $\bm{8}=\bm{3}_0 + \bm{2}_3 + \bm{2}_{-3} + \bm{1}_0$.
The background gauge fields have the VEV in the $\bm{3}_0$ component and its basis is given by
\begin{equation}
    \mathcal{T}_i %= T^i 
    =\left\{
    \frac{1}{2} \mqty(0 &  1 & 0 \\ 1 &  0 & 0 \\ 0 & 0 & 0 ),
    \ 
    \frac{1}{2} \mqty(0 & -i & 0 \\ i &  0 & 0 \\ 0 & 0 & 0 ),
    \ 
    \frac{1}{2} \mqty(1 &  0 & 0 \\ 0 & -1 & 0 \\ 0 & 0 & 0 )\right\}.
\end{equation}
On the other hand, the $\bm{2}_3 + \bm{2}_{-3}$ components are spanned by
\begin{align}
    E^{(+,\uparrow)}
    &=
    \frac{1}{\sqrt{2}}\mqty(0 & 0 & 1 \\ 0 & 0 & 0 \\  0 & 0 & 0 ),
    \\
    E^{(+,\downarrow)}
    &=
    \frac{1}{\sqrt{2}}\mqty(0 & 0 & 0 \\ 0 & 0 & 1 \\  0 & 0 & 0 ),
    \\
    E^{(-,\uparrow)}
    &=
    \frac{1}{\sqrt{2}}\mqty(0 & 0 & 0 \\ 0 & 0 & 0 \\  0 & 1 & 0 ),
    \\
    E^{(-,\downarrow)}
    &=
    \frac{1}{\sqrt{2}}\mqty(0 & 0 & 0 \\ 0 & 0 & 0 \\ -1 & 0 & 0 ),
\end{align}
where $q = +, -$ and $s = \uparrow, \downarrow$ of $E^{(q,s)}$ represent the sign of U($1$) charge and spin, respectively.
Note that these matrices satisfy
\begin{equation}
    \qty[i \mathcal{T}_i, E^{(q,s)}] = \frac{i \sigma_{i,s's}}{2} E^{(q,s')},
    \label{TE_commutator}
\end{equation}
where $\sigma_i$ are the Pauli matrices and $\sigma_{x,\uparrow \downarrow}$, for example, represents $\sigma_{x, 1 2}$.

The $\bm{1}_0$ component corresponds to the generator of the U($1$) subalgebra and its basis is represented by 
\begin{align}
    T_8
    &=
    \frac{1}{2\sqrt{3}}\mqty(-1 & 0 & 0 \\ 0 & -1 & 0 \\ 0 & 0 & 2 ).
\end{align}

With this decomposition of the adjoint representation,
we represent the gauge fields as 
\begin{align}
    A_i
    =
    A_i^j T_j + A_i^{(q,s)} E^{(q,s)} + A_i^8 T_8.
\end{align}
Note that $A_i^{(q,s)}$ satisfies
\begin{align}
    \qty(A_i^{(q,s)})^* = -\epsilon^{q q'} \epsilon^{s s'} A_i^{\qty(q',s')},
\end{align}
where $\epsilon^{q q'}$ is an antisymmetric tensor with $\epsilon^{+ -} = 1$.
In the following, we use the doublets
\begin{align}
    A_i^\pm
    \equiv
    \mqty( A_i^{(\pm,\uparrow)} \\ A_i^{(\pm,\downarrow)} )
\end{align}
%%

%%%%%%%%%%%%%%%%%%%%%%%%%%%%%%%%%%%%%%%%%%%%
\subsubsection{Quadratic action}
\label{subsubsec: L2 in (a)}
%%%%%%%%%%%%%%%%%%%%%%%%%%%%%%%%%%%%%%%%%%%%

After some calculations, we obtain the quadratic action of the gauge field perturbations for each component as
\begin{align}
    L_{2,\mathbf{3}}
    =
    &\frac{1}{2}
    \qty[
        (\delta A_i^{j \prime})^2 - (\partial_i \delta A_j^k)^2 
        + \partial_i \delta A_j^k \partial_j \delta A_i^k
    ]
    \nonumber\\
    &+ g a Q \epsilon^{i l k}
    \qty( \partial_i \delta A_j^k - \partial_j \delta A_i^k ) \delta A_j^l 
    \nonumber\\
    &-a \xi H \epsilon^{i j k}
    \qty(
        \delta A_i^l \partial_j \delta A_k^l 
        - g a Q \epsilon^{i l m} \delta A_j^l \delta A_k^m
    ).
\end{align}
\begin{align}
    L_{2,\mathbf{2}+\mathbf{2}}
    =
    &\qty(\delta A_i^{+\prime})^\dagger \delta A_i^{+\prime}
    -\qty(\partial_i \delta A_j^+)^\dagger \partial_i \delta A_j^+
    +\qty(\partial_i \delta A_j^+)^\dagger \partial_j \delta A_i^+
    \nonumber\\
    &+ g a Q \qty[
        \qty(\delta A_j^+)^\dagger \frac{i\sigma_i}{2} \partial_i \delta A_j^+
        -\qty(\partial_i \delta A_j^+)^\dagger \frac{i\sigma_i}{2} \delta A_j^+
        -\qty(\delta A_j^+)^\dagger \frac{i\sigma_i}{2} \partial_j \delta A_i^+
        +\qty(\partial_j \delta A_i^+)^\dagger \frac{i\sigma_i}{2} \delta A_j^+
    ]
    \nonumber\\
    &-\frac{\qty(g a Q)^2}{2} \qty[
        \qty(\delta A_i^+)^\dagger \delta A_i^+
        -3 \epsilon^{i j k} \qty(\delta A_i^+)^\dagger \frac{i\sigma_j}{2} \delta A_k^+
    ]
    \nonumber\\
    &- a\xi H \epsilon^{i j k}
    \qty[
        \qty(\delta A_i^+)^\dagger \partial_j \delta A_k^+
        - \partial_j \qty(\delta A_i^+)^\dagger \delta A_k^+
        + 2 g a Q  \qty(\delta A_i^+)^\dagger \frac{i\sigma_j}{2} \delta A_k^+
    ].
\end{align}
\begin{align}
    L_{2,\mathbf{1}}
    =
    &\frac{1}{2}
    \qty[
        (\delta A_i^{8 \prime})^2 - (\partial_i \delta A_j^8)^2 
        + \partial_i \delta A_j^8 \partial_j \delta A_i^8
    ]
    - a \xi H \epsilon^{i j k} \delta A_i^8 \partial_j \delta A_k^8,
\end{align}
where prime denotes the derivative with respect to $\tau$.
Extracting the terms with the traceless and transverse components $t_{i j}$ of $\delta A_i^j$ in $L_{2,\mathbf{3}}$,
we obtain
\begin{align}
    L_{2,\bm{3}}^{\mathrm{TT}}
    &=
    \frac{1}{2}
    \qty[
        t_{i j}' t_{i j}' 
        - \partial_i t_{j k} \partial_i t_{j k}
    ]
    + \frac{\xi + m_{Q+}}{\tau} 
    \epsilon^{i j k} t_{i l} \partial_j t_{k l} 
    - \frac{\xi m_{Q+}}{\tau^2} t_{i j} t_{i j}
    ,
\end{align}
where we used $\tau \simeq -1/(a H)$ during inflation.
Except for the replacement of $g$ with $g \lambda$, these actions are the same as in the CNI model~\cite{Adshead:2013nka}.

Note that the $\bm{3}$ and $\bm{1}$ components have the same action as that for the SU(2) and U(1) gauge fields, respectively.
On the other hand, the action of the $\bm{2} + \bm{2}$ component is unique to case (a).
By assuming the transverse condition, the action of the $\bm{2}+\bm{2}$ component becomes
\begin{align}
    L_{2,\mathbf{2}+\mathbf{2}}
    =&
    \qty(\delta A_i^{+\prime})^\dagger \delta A_i^{+\prime}
    \nonumber\\
    &+
    \qty(\delta A_i^+)^\dagger
    \qty[
        \delta_{i k}
        \qty(
            \partial^2
            +2 g a Q \frac{i \sigma_l}{2} \partial_l
            - \frac{\qty(g a Q)^2}{2}
        )
        + \epsilon^{i j k}
        \qty(
            \frac{3 \qty(g a Q)^2}{2} \frac{i\sigma_j}{2}
            - 2 a \xi H 
            \qty(\partial_j + g a Q \frac{i\sigma_j}{2})
        )
    ]
    \delta A_k^+
\end{align}
Considering the Fourier modes of the gauge field perturbations, the dynamics of these components can be described by the ``mass matrix'' given by
\begin{align}
    \mathcal{M}^2
    =
    \delta_{i k}
    \qty(
        k^2
        + g a Q \sigma_l k_l
        + \frac{\qty(g a Q)^2}{2}
    )
    + i \epsilon^{i j k}
    \qty(
        -\frac{3 \qty(g a Q)^2}{4}\sigma_j
        + 2 a \xi H 
        \qty(k_j + \frac{g a Q}{2} \sigma_j)
    ).
\end{align}
%%
%%%%%%%%%%%%%%%%%%%%%%%%%%%%%%%%%%%%%%%%%%%%
\subsection{Case (b)}
\label{subsec: case b}
%%%%%%%%%%%%%%%%%%%%%%%%%%%%%%%%%%%%%%%%%%%%

Next, we discuss the perturbations around the background solution in case (b).
Note that $\lambda = 1/2$ in this case.

%%%%%%%%%%%%%%%%%%%%%%%%%%%%%%%%%%%%%%%%%%%%
\subsubsection{Decomposition of the adjoint representation}
\label{subsubsec: decomposition in (b)}
%%%%%%%%%%%%%%%%%%%%%%%%%%%%%%%%%%%%%%%%%%%%

In this case, we consider the embeddings of SU(3)$\supset$SU(2) and then the adjoint representation of SU(3) is decomposed as $\bm{8}=\bm{3}+\bm{5}$.
The background gauge fields have the VEV in $\bm{3}$ component and its basis is given by
\begin{equation}
    \mathcal{T}_{i}
    =
    \left\{
        \frac{1}{2\sqrt{2}} \mqty(0 &  1 & 0 \\ 1 & 0 &  1 \\ 0 & 1 &  0 ),\ 
        \frac{1}{2\sqrt{2}} \mqty(0 & -i & 0 \\ i & 0 & -i \\ 0 & i &  0 ),\ 
        \frac{1}{2}         \mqty(1 &  0 & 0 \\ 0 & 0 &  0 \\ 0 & 0 & -1 )
    \right\}
\end{equation}
Note that the SU($2$) subalgebra is generated by $\mathcal{T}_{i}/\lambda = 2 \mathcal{T}_{i}$.

We decompose the $\mathbf{5}$ representation by 
\begin{align}
    \mathcal{T}^{(p)}
    &=
    \left\{
        \frac{1}{2}\mqty(0 & 0 & -i \\ 0 & 0 & 0 \\ i & 0 & 0 ),
        \quad
        \frac{1}{2\sqrt{2}}\mqty(0 & 1 & 0 \\ 1 & 0 & -1 \\  0 & -1 & 0 ),
        \quad
        \frac{1}{2\sqrt{3}}\mqty(-1 & 0 & 0 \\ 0 & 2 & 0 \\  0 & 0 & -1 ),
        \quad
        \frac{1}{2\sqrt{2}}\mqty(0 & -i & 0 \\ i & 0 & i \\ 0 & -i & 0 ),
        \quad
        \frac{1}{2}\mqty(0 & 0 & 1 \\ 0 & 0 & 0 \\  1 & 0 & 0 )
    \right\}.
\end{align}
Using this basis, the action of the SU($2$) generators are 
\begin{equation}
    \qty[i \mathcal{T}_i, \mathcal{T}^{(q)}] 
    =
    \frac{i\tilde{R}_i^{(p)(q)}}{2} \mathcal{T}^{(p)},
    \label{TT_commutator}
\end{equation}
with
\begin{align}
    i\tilde{R}_i = 
    \left\{
    \mqty(
        0 & 1 & 0 & 0 & 0
        \\
        -1 & 0 & 0 & 0 & 0
        \\
        0 & 0 & 0 & \sqrt{3} & 0
        \\
        0 & 0 & -\sqrt{3} & 0 & 1
        \\
        0 & 0 & 0 & -1 & 0
    ), \quad
    \mqty(
        0 & 0 & 0 & -1 & 0
        \\
        0 & 0 & \sqrt{3} & 0 & 1
        \\
        0 & -\sqrt{3} & 0 & 0 & 0
        \\
        1 & 0 & 0 & 0 & 0
        \\
        0 & -1 & 0 & 0 & 0
    ), \quad
    \mqty(
        0 & 0 & 0 & 0 & -2
        \\
        0 & 0 & 0 & 1 & 0
        \\
        0 & 0 & 0 & 0 & 0
        \\
        0 & -1 & 0 & 0 & 0
        \\
        2 & 0 & 0 & 0 & 0
    )
    \right\}.
\end{align}

With this decomposition of the adjoint representation,
we represent the gauge fields as 
\begin{align}
    A_i
    =
    A_i^{j} \mathcal{T}_{j} + A_i^{(p)} \mathcal{T}^{(p)}.
\end{align}
In the following, we use the vector
\begin{align}
    A_i^{\bm{5}}
    \equiv
    \mqty( 
        A_i^{(1)} \\ 
        A_i^{(2)} \\
        A_i^{(3)} \\ 
        A_i^{(4)} \\ 
        A_i^{(5)}  
    ).
\end{align}
%%

%%%%%%%%%%%%%%%%%%%%%%%%%%%%%%%%%%%%%%%%%%%%
\subsubsection{Quadratic action}
\label{subsubsec: L2 in (b)}
%%%%%%%%%%%%%%%%%%%%%%%%%%%%%%%%%%%%%%%%%%%%

After some calculations, we obtain the quadratic action of the gauge field perturbations for each component as
\begin{align}
    L^{(b)}_{2,\mathbf{3}}
    =
    &\frac{1}{2}
    \qty[
        \qty(\delta A_i^{j \prime})^2 
        - \qty(\partial_i \delta A_j^{k})^2 
        + \partial_i \delta A_j^{k} \partial_j \delta A_i^{k}
    ]
    \nonumber\\
    &+ g a Q \lambda \epsilon^{i l k}
    \qty( \partial_i \delta A_j^{k} - \partial_j \delta A_i^{k} ) \delta A_j^{l} 
    \nonumber\\
    &-a \xi H \epsilon^{i j k}
    \qty(
        \delta A_i^{l} \partial_j \delta A_k^{l} 
        - g a Q \lambda \epsilon^{i l m} \delta A_j^{l} \delta A_k^{m}
    ).
\end{align}
This is the same as Case (a) except for the replacement $g \to g \lambda$.

On the other hand, the $\bm{5}$ component has a unique action as in case (a):
\begin{align}
    L^{(b)}_{2,\mathbf{5}}
    =
    &\frac{1}{2}
    \qty[
        \qty(\delta A_i^{\mathbf{5} \prime})^\mathsf{T} \delta A_i^{\mathbf{5} \prime}
        - \qty(\partial_i \delta A_j^{\mathbf{5}})^\mathsf{T}
            \partial_i \delta A_j^{\mathbf{5}}
        + \qty(\partial_i \delta A_j^{\mathbf{5}})^\mathsf{T}
            \partial_j \delta A_i^{\mathbf{5}}
    ]
    \nonumber\\
    &- g a Q \lambda 
    \qty( \partial_i \delta A_j^{\mathbf{5}} - \partial_j \delta A_i^{\mathbf{5}} )^\mathsf{T} i\tilde{R}_i \delta A_j^{\mathbf{5}} 
    \nonumber\\
    &-\frac{\qty(g a Q \lambda)^2}{2} 
    \qty[
        6 \qty(\delta A_i^\mathbf{5})^{\mathsf{T}} \delta A_i^{\mathbf{5}}
        - \qty(i\tilde{R}_i \delta A_j^{\mathbf{5}})^{\mathsf{T}}
            i\tilde{R}_j \delta A_i^{\mathbf{5}}
        - \epsilon^{i j k} \qty(\delta A_i^{\mathbf{5}})^{\mathsf{T}}
            i\tilde{R}_j \delta A_k^{\mathbf{5}}
    ]
    \nonumber\\
    &-a \xi H \epsilon^{i j k}
    \qty[
        \qty(\delta A_i^{\mathbf{5}})^\mathsf{T}
            \partial_j \delta A_k^{\mathbf{5}}
        + g a Q \lambda \qty(\delta A_i^{\mathbf{5}})^{\mathsf{T}}
            i\tilde{R}_j \delta A_k^{\mathbf{5}}
    ].
\end{align}
By assuming the transverse condition, the action of the $\bm{5}$ component becomes
\begin{align}
    L^{(b)}_{2,\mathbf{5}}
    =&
    \frac{1}{2}
    \qty(\delta A_i^{\mathbf{5} \prime})^\mathsf{T} \delta A_i^{\mathbf{5} \prime}
    \nonumber\\
    &+
    \qty(\delta A_i^{\mathbf{5}})^\mathsf{T}
    \left[
        \delta_{i k}
        \qty(
            \frac{\partial^2}{2}
            + g a Q \lambda i\tilde{R}_j \partial_j
            -3 \qty(g a Q \lambda)^2
        )
    \right.
    \nonumber\\
    & \qquad \qquad \quad
    - \frac{\qty(g a Q \lambda)^2}{2} \tilde{R}_k^\mathsf{T} \tilde{R}_i
    +
    \left.
        \epsilon^{i j k}
        \qty{
            \qty(
                \frac{g a Q \lambda}{2} - a \xi H
            ) g a Q \lambda i\tilde{R}_j
            -a \xi H \partial_j 
        }
    \right]
    \delta A_k^{\mathbf{5}},
\end{align}
and then the ``mass matrix'' is 
\begin{align}
    \mathcal{M}^2
    =
    \delta_{i k}
    \qty[
        k^2
        + 2 g a Q \lambda \tilde{R}_l k_l
        + 6 \qty(g a Q \lambda)^2
    ]
    + \qty(g a Q \lambda)^2 \tilde{R}_k^\mathsf{T} \tilde{R}_i
    - \epsilon^{i j k}
    \qty[
        \qty(
            g a Q \lambda - 2 a \xi H
        ) g a Q \lambda i\tilde{R}_j
        - 2i a \xi H k_j 
    ].
\end{align}
%%

%%%%%%%%%%% References %%%%%%%%%%%
\small
\bibliographystyle{apsrev4-2}
\bibliography{Ref}

\begin{thebibliography}{105}%
\makeatletter
\providecommand \@ifxundefined [1]{%
 \@ifx{#1\undefined}
}%
\providecommand \@ifnum [1]{%
 \ifnum #1\expandafter \@firstoftwo
 \else \expandafter \@secondoftwo
 \fi
}%
\providecommand \@ifx [1]{%
 \ifx #1\expandafter \@firstoftwo
 \else \expandafter \@secondoftwo
 \fi
}%
\providecommand \natexlab [1]{#1}%
\providecommand \enquote  [1]{``#1''}%
\providecommand \bibnamefont  [1]{#1}%
\providecommand \bibfnamefont [1]{#1}%
\providecommand \citenamefont [1]{#1}%
\providecommand \href@noop [0]{\@secondoftwo}%
\providecommand \href [0]{\begingroup \@sanitize@url \@href}%
\providecommand \@href[1]{\@@startlink{#1}\@@href}%
\providecommand \@@href[1]{\endgroup#1\@@endlink}%
\providecommand \@sanitize@url [0]{\catcode `\\12\catcode `\$12\catcode
  `\&12\catcode `\#12\catcode `\^12\catcode `\_12\catcode `\%12\relax}%
\providecommand \@@startlink[1]{}%
\providecommand \@@endlink[0]{}%
\providecommand \url  [0]{\begingroup\@sanitize@url \@url }%
\providecommand \@url [1]{\endgroup\@href {#1}{\urlprefix }}%
\providecommand \urlprefix  [0]{URL }%
\providecommand \Eprint [0]{\href }%
\providecommand \doibase [0]{https://doi.org/}%
\providecommand \selectlanguage [0]{\@gobble}%
\providecommand \bibinfo  [0]{\@secondoftwo}%
\providecommand \bibfield  [0]{\@secondoftwo}%
\providecommand \translation [1]{[#1]}%
\providecommand \BibitemOpen [0]{}%
\providecommand \bibitemStop [0]{}%
\providecommand \bibitemNoStop [0]{.\EOS\space}%
\providecommand \EOS [0]{\spacefactor3000\relax}%
\providecommand \BibitemShut  [1]{\csname bibitem#1\endcsname}%
\let\auto@bib@innerbib\@empty
%</preamble>
\bibitem [{\citenamefont {Freese}\ \emph {et~al.}(1990)\citenamefont {Freese},
  \citenamefont {Frieman},\ and\ \citenamefont {Olinto}}]{Freese:1990rb}%
  \BibitemOpen
  \bibfield  {author} {\bibinfo {author} {\bibfnamefont {K.}~\bibnamefont
  {Freese}}, \bibinfo {author} {\bibfnamefont {J.~A.}\ \bibnamefont
  {Frieman}},\ and\ \bibinfo {author} {\bibfnamefont {A.~V.}\ \bibnamefont
  {Olinto}},\ }\href {https://doi.org/10.1103/PhysRevLett.65.3233} {\bibfield
  {journal} {\bibinfo  {journal} {Phys. Rev. Lett.}\ }\textbf {\bibinfo
  {volume} {65}},\ \bibinfo {pages} {3233} (\bibinfo {year}
  {1990})}\BibitemShut {NoStop}%
\bibitem [{\citenamefont {Adams}\ \emph {et~al.}(1993)\citenamefont {Adams},
  \citenamefont {Bond}, \citenamefont {Freese}, \citenamefont {Frieman},\ and\
  \citenamefont {Olinto}}]{Adams:1992bn}%
  \BibitemOpen
  \bibfield  {author} {\bibinfo {author} {\bibfnamefont {F.~C.}\ \bibnamefont
  {Adams}}, \bibinfo {author} {\bibfnamefont {J.~R.}\ \bibnamefont {Bond}},
  \bibinfo {author} {\bibfnamefont {K.}~\bibnamefont {Freese}}, \bibinfo
  {author} {\bibfnamefont {J.~A.}\ \bibnamefont {Frieman}},\ and\ \bibinfo
  {author} {\bibfnamefont {A.~V.}\ \bibnamefont {Olinto}},\ }\href
  {https://doi.org/10.1103/PhysRevD.47.426} {\bibfield  {journal} {\bibinfo
  {journal} {Phys. Rev. D}\ }\textbf {\bibinfo {volume} {47}},\ \bibinfo
  {pages} {426} (\bibinfo {year} {1993})},\ \Eprint
  {https://arxiv.org/abs/hep-ph/9207245} {arXiv:hep-ph/9207245} \BibitemShut
  {NoStop}%
\bibitem [{\citenamefont {Kim}\ \emph {et~al.}(2005)\citenamefont {Kim},
  \citenamefont {Nilles},\ and\ \citenamefont {Peloso}}]{Kim:2004rp}%
  \BibitemOpen
  \bibfield  {author} {\bibinfo {author} {\bibfnamefont {J.~E.}\ \bibnamefont
  {Kim}}, \bibinfo {author} {\bibfnamefont {H.~P.}\ \bibnamefont {Nilles}},\
  and\ \bibinfo {author} {\bibfnamefont {M.}~\bibnamefont {Peloso}},\ }\href
  {https://doi.org/10.1088/1475-7516/2005/01/005} {\bibfield  {journal}
  {\bibinfo  {journal} {JCAP}\ }\textbf {\bibinfo {volume} {01}},\ \bibinfo
  {pages} {005}},\ \Eprint {https://arxiv.org/abs/hep-ph/0409138}
  {arXiv:hep-ph/0409138} \BibitemShut {NoStop}%
\bibitem [{\citenamefont {Freese}\ and\ \citenamefont
  {Kinney}(2004)}]{Freese:2004un}%
  \BibitemOpen
  \bibfield  {author} {\bibinfo {author} {\bibfnamefont {K.}~\bibnamefont
  {Freese}}\ and\ \bibinfo {author} {\bibfnamefont {W.~H.}\ \bibnamefont
  {Kinney}},\ }\href {https://doi.org/10.1103/PhysRevD.70.083512} {\bibfield
  {journal} {\bibinfo  {journal} {Phys. Rev. D}\ }\textbf {\bibinfo {volume}
  {70}},\ \bibinfo {pages} {083512} (\bibinfo {year} {2004})},\ \Eprint
  {https://arxiv.org/abs/hep-ph/0404012} {arXiv:hep-ph/0404012} \BibitemShut
  {NoStop}%
\bibitem [{\citenamefont {Arkani-Hamed}\ \emph {et~al.}(2003)\citenamefont
  {Arkani-Hamed}, \citenamefont {Cheng}, \citenamefont {Creminelli},\ and\
  \citenamefont {Randall}}]{Arkani-Hamed:2003xts}%
  \BibitemOpen
  \bibfield  {author} {\bibinfo {author} {\bibfnamefont {N.}~\bibnamefont
  {Arkani-Hamed}}, \bibinfo {author} {\bibfnamefont {H.-C.}\ \bibnamefont
  {Cheng}}, \bibinfo {author} {\bibfnamefont {P.}~\bibnamefont {Creminelli}},\
  and\ \bibinfo {author} {\bibfnamefont {L.}~\bibnamefont {Randall}},\ }\href
  {https://doi.org/10.1103/PhysRevLett.90.221302} {\bibfield  {journal}
  {\bibinfo  {journal} {Phys. Rev. Lett.}\ }\textbf {\bibinfo {volume} {90}},\
  \bibinfo {pages} {221302} (\bibinfo {year} {2003})},\ \Eprint
  {https://arxiv.org/abs/hep-th/0301218} {arXiv:hep-th/0301218} \BibitemShut
  {NoStop}%
\bibitem [{\citenamefont {Banks}\ \emph {et~al.}(2003)\citenamefont {Banks},
  \citenamefont {Dine}, \citenamefont {Fox},\ and\ \citenamefont
  {Gorbatov}}]{Banks:2003sx}%
  \BibitemOpen
  \bibfield  {author} {\bibinfo {author} {\bibfnamefont {T.}~\bibnamefont
  {Banks}}, \bibinfo {author} {\bibfnamefont {M.}~\bibnamefont {Dine}},
  \bibinfo {author} {\bibfnamefont {P.~J.}\ \bibnamefont {Fox}},\ and\ \bibinfo
  {author} {\bibfnamefont {E.}~\bibnamefont {Gorbatov}},\ }\href
  {https://doi.org/10.1088/1475-7516/2003/06/001} {\bibfield  {journal}
  {\bibinfo  {journal} {JCAP}\ }\textbf {\bibinfo {volume} {06}},\ \bibinfo
  {pages} {001}},\ \Eprint {https://arxiv.org/abs/hep-th/0303252}
  {arXiv:hep-th/0303252} \BibitemShut {NoStop}%
\bibitem [{\citenamefont {Dimopoulos}\ \emph {et~al.}(2008)\citenamefont
  {Dimopoulos}, \citenamefont {Kachru}, \citenamefont {McGreevy},\ and\
  \citenamefont {Wacker}}]{Dimopoulos:2005ac}%
  \BibitemOpen
  \bibfield  {author} {\bibinfo {author} {\bibfnamefont {S.}~\bibnamefont
  {Dimopoulos}}, \bibinfo {author} {\bibfnamefont {S.}~\bibnamefont {Kachru}},
  \bibinfo {author} {\bibfnamefont {J.}~\bibnamefont {McGreevy}},\ and\
  \bibinfo {author} {\bibfnamefont {J.~G.}\ \bibnamefont {Wacker}},\ }\href
  {https://doi.org/10.1088/1475-7516/2008/08/003} {\bibfield  {journal}
  {\bibinfo  {journal} {JCAP}\ }\textbf {\bibinfo {volume} {08}},\ \bibinfo
  {pages} {003}},\ \Eprint {https://arxiv.org/abs/hep-th/0507205}
  {arXiv:hep-th/0507205} \BibitemShut {NoStop}%
\bibitem [{\citenamefont {Easther}\ and\ \citenamefont
  {McAllister}(2006)}]{Easther:2005zr}%
  \BibitemOpen
  \bibfield  {author} {\bibinfo {author} {\bibfnamefont {R.}~\bibnamefont
  {Easther}}\ and\ \bibinfo {author} {\bibfnamefont {L.}~\bibnamefont
  {McAllister}},\ }\href {https://doi.org/10.1088/1475-7516/2006/05/018}
  {\bibfield  {journal} {\bibinfo  {journal} {JCAP}\ }\textbf {\bibinfo
  {volume} {05}},\ \bibinfo {pages} {018}},\ \Eprint
  {https://arxiv.org/abs/hep-th/0512102} {arXiv:hep-th/0512102} \BibitemShut
  {NoStop}%
\bibitem [{\citenamefont {Kaloper}\ and\ \citenamefont
  {Sorbo}(2009)}]{Kaloper:2008fb}%
  \BibitemOpen
  \bibfield  {author} {\bibinfo {author} {\bibfnamefont {N.}~\bibnamefont
  {Kaloper}}\ and\ \bibinfo {author} {\bibfnamefont {L.}~\bibnamefont
  {Sorbo}},\ }\href {https://doi.org/10.1103/PhysRevLett.102.121301} {\bibfield
   {journal} {\bibinfo  {journal} {Phys. Rev. Lett.}\ }\textbf {\bibinfo
  {volume} {102}},\ \bibinfo {pages} {121301} (\bibinfo {year} {2009})},\
  \Eprint {https://arxiv.org/abs/0811.1989} {arXiv:0811.1989 [hep-th]}
  \BibitemShut {NoStop}%
\bibitem [{\citenamefont {Silverstein}\ and\ \citenamefont
  {Westphal}(2008)}]{Silverstein:2008sg}%
  \BibitemOpen
  \bibfield  {author} {\bibinfo {author} {\bibfnamefont {E.}~\bibnamefont
  {Silverstein}}\ and\ \bibinfo {author} {\bibfnamefont {A.}~\bibnamefont
  {Westphal}},\ }\href {https://doi.org/10.1103/PhysRevD.78.106003} {\bibfield
  {journal} {\bibinfo  {journal} {Phys. Rev. D}\ }\textbf {\bibinfo {volume}
  {78}},\ \bibinfo {pages} {106003} (\bibinfo {year} {2008})},\ \Eprint
  {https://arxiv.org/abs/0803.3085} {arXiv:0803.3085 [hep-th]} \BibitemShut
  {NoStop}%
\bibitem [{\citenamefont {McAllister}\ \emph {et~al.}(2010)\citenamefont
  {McAllister}, \citenamefont {Silverstein},\ and\ \citenamefont
  {Westphal}}]{McAllister:2008hb}%
  \BibitemOpen
  \bibfield  {author} {\bibinfo {author} {\bibfnamefont {L.}~\bibnamefont
  {McAllister}}, \bibinfo {author} {\bibfnamefont {E.}~\bibnamefont
  {Silverstein}},\ and\ \bibinfo {author} {\bibfnamefont {A.}~\bibnamefont
  {Westphal}},\ }\href {https://doi.org/10.1103/PhysRevD.82.046003} {\bibfield
  {journal} {\bibinfo  {journal} {Phys. Rev. D}\ }\textbf {\bibinfo {volume}
  {82}},\ \bibinfo {pages} {046003} (\bibinfo {year} {2010})},\ \Eprint
  {https://arxiv.org/abs/0808.0706} {arXiv:0808.0706 [hep-th]} \BibitemShut
  {NoStop}%
\bibitem [{\citenamefont {Flauger}\ \emph {et~al.}(2010)\citenamefont
  {Flauger}, \citenamefont {McAllister}, \citenamefont {Pajer}, \citenamefont
  {Westphal},\ and\ \citenamefont {Xu}}]{Flauger:2009ab}%
  \BibitemOpen
  \bibfield  {author} {\bibinfo {author} {\bibfnamefont {R.}~\bibnamefont
  {Flauger}}, \bibinfo {author} {\bibfnamefont {L.}~\bibnamefont {McAllister}},
  \bibinfo {author} {\bibfnamefont {E.}~\bibnamefont {Pajer}}, \bibinfo
  {author} {\bibfnamefont {A.}~\bibnamefont {Westphal}},\ and\ \bibinfo
  {author} {\bibfnamefont {G.}~\bibnamefont {Xu}},\ }\href
  {https://doi.org/10.1088/1475-7516/2010/06/009} {\bibfield  {journal}
  {\bibinfo  {journal} {JCAP}\ }\textbf {\bibinfo {volume} {06}},\ \bibinfo
  {pages} {009}},\ \Eprint {https://arxiv.org/abs/0907.2916} {arXiv:0907.2916
  [hep-th]} \BibitemShut {NoStop}%
\bibitem [{\citenamefont {Anber}\ and\ \citenamefont
  {Sorbo}(2010)}]{Anber:2009ua}%
  \BibitemOpen
  \bibfield  {author} {\bibinfo {author} {\bibfnamefont {M.~M.}\ \bibnamefont
  {Anber}}\ and\ \bibinfo {author} {\bibfnamefont {L.}~\bibnamefont {Sorbo}},\
  }\href {https://doi.org/10.1103/PhysRevD.81.043534} {\bibfield  {journal}
  {\bibinfo  {journal} {Phys. Rev. D}\ }\textbf {\bibinfo {volume} {81}},\
  \bibinfo {pages} {043534} (\bibinfo {year} {2010})},\ \Eprint
  {https://arxiv.org/abs/0908.4089} {arXiv:0908.4089 [hep-th]} \BibitemShut
  {NoStop}%
\bibitem [{\citenamefont {Akrami}\ \emph
  {et~al.}(2020{\natexlab{a}})\citenamefont {Akrami} \emph
  {et~al.}}]{Planck:2018jri}%
  \BibitemOpen
  \bibfield  {author} {\bibinfo {author} {\bibfnamefont {Y.}~\bibnamefont
  {Akrami}} \emph {et~al.} (\bibinfo {collaboration} {Planck}),\ }\href
  {https://doi.org/10.1051/0004-6361/201833887} {\bibfield  {journal} {\bibinfo
   {journal} {Astron. Astrophys.}\ }\textbf {\bibinfo {volume} {641}},\
  \bibinfo {pages} {A10} (\bibinfo {year} {2020}{\natexlab{a}})},\ \Eprint
  {https://arxiv.org/abs/1807.06211} {arXiv:1807.06211 [astro-ph.CO]}
  \BibitemShut {NoStop}%
\bibitem [{\citenamefont {Peloso}\ and\ \citenamefont
  {Unal}(2015)}]{Peloso:2015dsa}%
  \BibitemOpen
  \bibfield  {author} {\bibinfo {author} {\bibfnamefont {M.}~\bibnamefont
  {Peloso}}\ and\ \bibinfo {author} {\bibfnamefont {C.}~\bibnamefont {Unal}},\
  }\href {https://doi.org/10.1088/1475-7516/2015/06/040} {\bibfield  {journal}
  {\bibinfo  {journal} {JCAP}\ }\textbf {\bibinfo {volume} {06}},\ \bibinfo
  {pages} {040}},\ \Eprint {https://arxiv.org/abs/1504.02784} {arXiv:1504.02784
  [astro-ph.CO]} \BibitemShut {NoStop}%
\bibitem [{\citenamefont {Notari}\ and\ \citenamefont
  {Tywoniuk}(2016)}]{Notari:2016npn}%
  \BibitemOpen
  \bibfield  {author} {\bibinfo {author} {\bibfnamefont {A.}~\bibnamefont
  {Notari}}\ and\ \bibinfo {author} {\bibfnamefont {K.}~\bibnamefont
  {Tywoniuk}},\ }\href {https://doi.org/10.1088/1475-7516/2016/12/038}
  {\bibfield  {journal} {\bibinfo  {journal} {JCAP}\ }\textbf {\bibinfo
  {volume} {12}},\ \bibinfo {pages} {038}},\ \Eprint
  {https://arxiv.org/abs/1608.06223} {arXiv:1608.06223 [hep-th]} \BibitemShut
  {NoStop}%
\bibitem [{\citenamefont {Ferreira}\ and\ \citenamefont
  {Notari}(2017)}]{Ferreira:2017lnd}%
  \BibitemOpen
  \bibfield  {author} {\bibinfo {author} {\bibfnamefont {R.~Z.}\ \bibnamefont
  {Ferreira}}\ and\ \bibinfo {author} {\bibfnamefont {A.}~\bibnamefont
  {Notari}},\ }\href {https://doi.org/10.1088/1475-7516/2017/09/007} {\bibfield
   {journal} {\bibinfo  {journal} {JCAP}\ }\textbf {\bibinfo {volume} {09}},\
  \bibinfo {pages} {007}},\ \Eprint {https://arxiv.org/abs/1706.00373}
  {arXiv:1706.00373 [astro-ph.CO]} \BibitemShut {NoStop}%
\bibitem [{\citenamefont {Tangarife}\ \emph {et~al.}(2017)\citenamefont
  {Tangarife}, \citenamefont {Tobioka}, \citenamefont {Ubaldi},\ and\
  \citenamefont {Volansky}}]{Tangarife:2017vnd}%
  \BibitemOpen
  \bibfield  {author} {\bibinfo {author} {\bibfnamefont {W.}~\bibnamefont
  {Tangarife}}, \bibinfo {author} {\bibfnamefont {K.}~\bibnamefont {Tobioka}},
  \bibinfo {author} {\bibfnamefont {L.}~\bibnamefont {Ubaldi}},\ and\ \bibinfo
  {author} {\bibfnamefont {T.}~\bibnamefont {Volansky}},\ }\href@noop {} {\
  (\bibinfo {year} {2017})},\ \Eprint {https://arxiv.org/abs/1706.00438}
  {arXiv:1706.00438 [hep-ph]} \BibitemShut {NoStop}%
\bibitem [{\citenamefont {Tangarife}\ \emph {et~al.}(2018)\citenamefont
  {Tangarife}, \citenamefont {Tobioka}, \citenamefont {Ubaldi},\ and\
  \citenamefont {Volansky}}]{Tangarife:2017rgl}%
  \BibitemOpen
  \bibfield  {author} {\bibinfo {author} {\bibfnamefont {W.}~\bibnamefont
  {Tangarife}}, \bibinfo {author} {\bibfnamefont {K.}~\bibnamefont {Tobioka}},
  \bibinfo {author} {\bibfnamefont {L.}~\bibnamefont {Ubaldi}},\ and\ \bibinfo
  {author} {\bibfnamefont {T.}~\bibnamefont {Volansky}},\ }\href
  {https://doi.org/10.1007/JHEP02(2018)084} {\bibfield  {journal} {\bibinfo
  {journal} {JHEP}\ }\textbf {\bibinfo {volume} {02}},\ \bibinfo {pages}
  {084}},\ \Eprint {https://arxiv.org/abs/1706.03072} {arXiv:1706.03072
  [hep-ph]} \BibitemShut {NoStop}%
\bibitem [{\citenamefont {Beltr\'an~Almeida}\ and\ \citenamefont
  {Bernal}(2018)}]{Almeida:2018pir}%
  \BibitemOpen
  \bibfield  {author} {\bibinfo {author} {\bibfnamefont {J.~P.}\ \bibnamefont
  {Beltr\'an~Almeida}}\ and\ \bibinfo {author} {\bibfnamefont {N.}~\bibnamefont
  {Bernal}},\ }\href {https://doi.org/10.1103/PhysRevD.98.083519} {\bibfield
  {journal} {\bibinfo  {journal} {Phys. Rev. D}\ }\textbf {\bibinfo {volume}
  {98}},\ \bibinfo {pages} {083519} (\bibinfo {year} {2018})},\ \Eprint
  {https://arxiv.org/abs/1803.09743} {arXiv:1803.09743 [astro-ph.CO]}
  \BibitemShut {NoStop}%
\bibitem [{\citenamefont {Lue}\ \emph {et~al.}(1999)\citenamefont {Lue},
  \citenamefont {Wang},\ and\ \citenamefont {Kamionkowski}}]{Lue:1998mq}%
  \BibitemOpen
  \bibfield  {author} {\bibinfo {author} {\bibfnamefont {A.}~\bibnamefont
  {Lue}}, \bibinfo {author} {\bibfnamefont {L.-M.}\ \bibnamefont {Wang}},\ and\
  \bibinfo {author} {\bibfnamefont {M.}~\bibnamefont {Kamionkowski}},\ }\href
  {https://doi.org/10.1103/PhysRevLett.83.1506} {\bibfield  {journal} {\bibinfo
   {journal} {Phys. Rev. Lett.}\ }\textbf {\bibinfo {volume} {83}},\ \bibinfo
  {pages} {1506} (\bibinfo {year} {1999})},\ \Eprint
  {https://arxiv.org/abs/astro-ph/9812088} {arXiv:astro-ph/9812088}
  \BibitemShut {NoStop}%
\bibitem [{\citenamefont {Barnaby}\ and\ \citenamefont
  {Peloso}(2011)}]{Barnaby:2010vf}%
  \BibitemOpen
  \bibfield  {author} {\bibinfo {author} {\bibfnamefont {N.}~\bibnamefont
  {Barnaby}}\ and\ \bibinfo {author} {\bibfnamefont {M.}~\bibnamefont
  {Peloso}},\ }\href {https://doi.org/10.1103/PhysRevLett.106.181301}
  {\bibfield  {journal} {\bibinfo  {journal} {Phys. Rev. Lett.}\ }\textbf
  {\bibinfo {volume} {106}},\ \bibinfo {pages} {181301} (\bibinfo {year}
  {2011})},\ \Eprint {https://arxiv.org/abs/1011.1500} {arXiv:1011.1500
  [hep-ph]} \BibitemShut {NoStop}%
\bibitem [{\citenamefont {Sorbo}(2011)}]{Sorbo:2011rz}%
  \BibitemOpen
  \bibfield  {author} {\bibinfo {author} {\bibfnamefont {L.}~\bibnamefont
  {Sorbo}},\ }\href {https://doi.org/10.1088/1475-7516/2011/06/003} {\bibfield
  {journal} {\bibinfo  {journal} {JCAP}\ }\textbf {\bibinfo {volume} {06}},\
  \bibinfo {pages} {003}},\ \Eprint {https://arxiv.org/abs/1101.1525}
  {arXiv:1101.1525 [astro-ph.CO]} \BibitemShut {NoStop}%
\bibitem [{\citenamefont {Barnaby}\ \emph {et~al.}(2011)\citenamefont
  {Barnaby}, \citenamefont {Namba},\ and\ \citenamefont
  {Peloso}}]{Barnaby:2011vw}%
  \BibitemOpen
  \bibfield  {author} {\bibinfo {author} {\bibfnamefont {N.}~\bibnamefont
  {Barnaby}}, \bibinfo {author} {\bibfnamefont {R.}~\bibnamefont {Namba}},\
  and\ \bibinfo {author} {\bibfnamefont {M.}~\bibnamefont {Peloso}},\ }\href
  {https://doi.org/10.1088/1475-7516/2011/04/009} {\bibfield  {journal}
  {\bibinfo  {journal} {JCAP}\ }\textbf {\bibinfo {volume} {04}},\ \bibinfo
  {pages} {009}},\ \Eprint {https://arxiv.org/abs/1102.4333} {arXiv:1102.4333
  [astro-ph.CO]} \BibitemShut {NoStop}%
\bibitem [{\citenamefont {Barnaby}\ and\ \citenamefont
  {Shandera}(2012)}]{Barnaby:2011pe}%
  \BibitemOpen
  \bibfield  {author} {\bibinfo {author} {\bibfnamefont {N.}~\bibnamefont
  {Barnaby}}\ and\ \bibinfo {author} {\bibfnamefont {S.}~\bibnamefont
  {Shandera}},\ }\href {https://doi.org/10.1088/1475-7516/2012/01/034}
  {\bibfield  {journal} {\bibinfo  {journal} {JCAP}\ }\textbf {\bibinfo
  {volume} {01}},\ \bibinfo {pages} {034}},\ \Eprint
  {https://arxiv.org/abs/1109.2985} {arXiv:1109.2985 [astro-ph.CO]}
  \BibitemShut {NoStop}%
\bibitem [{\citenamefont {Dimopoulos}\ and\ \citenamefont
  {Karciauskas}(2012)}]{Dimopoulos:2012av}%
  \BibitemOpen
  \bibfield  {author} {\bibinfo {author} {\bibfnamefont {K.}~\bibnamefont
  {Dimopoulos}}\ and\ \bibinfo {author} {\bibfnamefont {M.}~\bibnamefont
  {Karciauskas}},\ }\href {https://doi.org/10.1007/JHEP06(2012)040} {\bibfield
  {journal} {\bibinfo  {journal} {JHEP}\ }\textbf {\bibinfo {volume} {06}},\
  \bibinfo {pages} {040}},\ \Eprint {https://arxiv.org/abs/1203.0230}
  {arXiv:1203.0230 [hep-ph]} \BibitemShut {NoStop}%
\bibitem [{\citenamefont {Anber}\ and\ \citenamefont
  {Sorbo}(2012)}]{Anber:2012du}%
  \BibitemOpen
  \bibfield  {author} {\bibinfo {author} {\bibfnamefont {M.~M.}\ \bibnamefont
  {Anber}}\ and\ \bibinfo {author} {\bibfnamefont {L.}~\bibnamefont {Sorbo}},\
  }\href {https://doi.org/10.1103/PhysRevD.85.123537} {\bibfield  {journal}
  {\bibinfo  {journal} {Phys. Rev. D}\ }\textbf {\bibinfo {volume} {85}},\
  \bibinfo {pages} {123537} (\bibinfo {year} {2012})},\ \Eprint
  {https://arxiv.org/abs/1203.5849} {arXiv:1203.5849 [astro-ph.CO]}
  \BibitemShut {NoStop}%
\bibitem [{\citenamefont {Meerburg}\ and\ \citenamefont
  {Pajer}(2013)}]{Meerburg:2012id}%
  \BibitemOpen
  \bibfield  {author} {\bibinfo {author} {\bibfnamefont {P.~D.}\ \bibnamefont
  {Meerburg}}\ and\ \bibinfo {author} {\bibfnamefont {E.}~\bibnamefont
  {Pajer}},\ }\href {https://doi.org/10.1088/1475-7516/2013/02/017} {\bibfield
  {journal} {\bibinfo  {journal} {JCAP}\ }\textbf {\bibinfo {volume} {02}},\
  \bibinfo {pages} {017}},\ \Eprint {https://arxiv.org/abs/1203.6076}
  {arXiv:1203.6076 [astro-ph.CO]} \BibitemShut {NoStop}%
\bibitem [{\citenamefont {Linde}\ \emph {et~al.}(2013)\citenamefont {Linde},
  \citenamefont {Mooij},\ and\ \citenamefont {Pajer}}]{Linde:2012bt}%
  \BibitemOpen
  \bibfield  {author} {\bibinfo {author} {\bibfnamefont {A.}~\bibnamefont
  {Linde}}, \bibinfo {author} {\bibfnamefont {S.}~\bibnamefont {Mooij}},\ and\
  \bibinfo {author} {\bibfnamefont {E.}~\bibnamefont {Pajer}},\ }\href
  {https://doi.org/10.1103/PhysRevD.87.103506} {\bibfield  {journal} {\bibinfo
  {journal} {Phys. Rev. D}\ }\textbf {\bibinfo {volume} {87}},\ \bibinfo
  {pages} {103506} (\bibinfo {year} {2013})},\ \Eprint
  {https://arxiv.org/abs/1212.1693} {arXiv:1212.1693 [hep-th]} \BibitemShut
  {NoStop}%
\bibitem [{\citenamefont {Ferreira}\ and\ \citenamefont
  {Sloth}(2014)}]{Ferreira:2014zia}%
  \BibitemOpen
  \bibfield  {author} {\bibinfo {author} {\bibfnamefont {R.~Z.}\ \bibnamefont
  {Ferreira}}\ and\ \bibinfo {author} {\bibfnamefont {M.~S.}\ \bibnamefont
  {Sloth}},\ }\href {https://doi.org/10.1007/JHEP12(2014)139} {\bibfield
  {journal} {\bibinfo  {journal} {JHEP}\ }\textbf {\bibinfo {volume} {12}},\
  \bibinfo {pages} {139}},\ \Eprint {https://arxiv.org/abs/1409.5799}
  {arXiv:1409.5799 [hep-ph]} \BibitemShut {NoStop}%
\bibitem [{\citenamefont {Bartolo}\ \emph {et~al.}(2015)\citenamefont
  {Bartolo}, \citenamefont {Matarrese}, \citenamefont {Peloso},\ and\
  \citenamefont {Shiraishi}}]{Bartolo:2015dga}%
  \BibitemOpen
  \bibfield  {author} {\bibinfo {author} {\bibfnamefont {N.}~\bibnamefont
  {Bartolo}}, \bibinfo {author} {\bibfnamefont {S.}~\bibnamefont {Matarrese}},
  \bibinfo {author} {\bibfnamefont {M.}~\bibnamefont {Peloso}},\ and\ \bibinfo
  {author} {\bibfnamefont {M.}~\bibnamefont {Shiraishi}},\ }\href
  {https://doi.org/10.1088/1475-7516/2015/07/039} {\bibfield  {journal}
  {\bibinfo  {journal} {JCAP}\ }\textbf {\bibinfo {volume} {07}},\ \bibinfo
  {pages} {039}},\ \Eprint {https://arxiv.org/abs/1505.02193} {arXiv:1505.02193
  [astro-ph.CO]} \BibitemShut {NoStop}%
\bibitem [{\citenamefont {Ferreira}\ \emph {et~al.}(2016)\citenamefont
  {Ferreira}, \citenamefont {Ganc}, \citenamefont {Nore\~na},\ and\
  \citenamefont {Sloth}}]{Ferreira:2015omg}%
  \BibitemOpen
  \bibfield  {author} {\bibinfo {author} {\bibfnamefont {R.~Z.}\ \bibnamefont
  {Ferreira}}, \bibinfo {author} {\bibfnamefont {J.}~\bibnamefont {Ganc}},
  \bibinfo {author} {\bibfnamefont {J.}~\bibnamefont {Nore\~na}},\ and\
  \bibinfo {author} {\bibfnamefont {M.~S.}\ \bibnamefont {Sloth}},\ }\href
  {https://doi.org/10.1088/1475-7516/2016/04/039} {\bibfield  {journal}
  {\bibinfo  {journal} {JCAP}\ }\textbf {\bibinfo {volume} {04}},\ \bibinfo
  {pages} {039}},\ \bibinfo {note} {[Erratum: JCAP 10, E01 (2016)]},\ \Eprint
  {https://arxiv.org/abs/1512.06116} {arXiv:1512.06116 [astro-ph.CO]}
  \BibitemShut {NoStop}%
\bibitem [{\citenamefont {Peloso}\ \emph {et~al.}(2016)\citenamefont {Peloso},
  \citenamefont {Sorbo},\ and\ \citenamefont {Unal}}]{Peloso:2016gqs}%
  \BibitemOpen
  \bibfield  {author} {\bibinfo {author} {\bibfnamefont {M.}~\bibnamefont
  {Peloso}}, \bibinfo {author} {\bibfnamefont {L.}~\bibnamefont {Sorbo}},\ and\
  \bibinfo {author} {\bibfnamefont {C.}~\bibnamefont {Unal}},\ }\href
  {https://doi.org/10.1088/1475-7516/2016/09/001} {\bibfield  {journal}
  {\bibinfo  {journal} {JCAP}\ }\textbf {\bibinfo {volume} {09}},\ \bibinfo
  {pages} {001}},\ \Eprint {https://arxiv.org/abs/1606.00459} {arXiv:1606.00459
  [astro-ph.CO]} \BibitemShut {NoStop}%
\bibitem [{\citenamefont {Alexander}\ \emph {et~al.}(2017)\citenamefont
  {Alexander}, \citenamefont {McDonough},\ and\ \citenamefont
  {Sims}}]{Alexander:2017bxe}%
  \BibitemOpen
  \bibfield  {author} {\bibinfo {author} {\bibfnamefont {S.}~\bibnamefont
  {Alexander}}, \bibinfo {author} {\bibfnamefont {E.}~\bibnamefont
  {McDonough}},\ and\ \bibinfo {author} {\bibfnamefont {R.}~\bibnamefont
  {Sims}},\ }\href {https://doi.org/10.1103/PhysRevD.96.063506} {\bibfield
  {journal} {\bibinfo  {journal} {Phys. Rev. D}\ }\textbf {\bibinfo {volume}
  {96}},\ \bibinfo {pages} {063506} (\bibinfo {year} {2017})},\ \Eprint
  {https://arxiv.org/abs/1704.00838} {arXiv:1704.00838 [gr-qc]} \BibitemShut
  {NoStop}%
\bibitem [{\citenamefont {Domcke}\ and\ \citenamefont
  {Mukaida}(2018)}]{Domcke:2018eki}%
  \BibitemOpen
  \bibfield  {author} {\bibinfo {author} {\bibfnamefont {V.}~\bibnamefont
  {Domcke}}\ and\ \bibinfo {author} {\bibfnamefont {K.}~\bibnamefont
  {Mukaida}},\ }\href {https://doi.org/10.1088/1475-7516/2018/11/020}
  {\bibfield  {journal} {\bibinfo  {journal} {JCAP}\ }\textbf {\bibinfo
  {volume} {11}},\ \bibinfo {pages} {020}},\ \Eprint
  {https://arxiv.org/abs/1806.08769} {arXiv:1806.08769 [hep-ph]} \BibitemShut
  {NoStop}%
\bibitem [{\citenamefont {Beltr\'an~Almeida}\ \emph {et~al.}(2019)\citenamefont
  {Beltr\'an~Almeida}, \citenamefont {Motoa-Manzano},\ and\ \citenamefont
  {Valenzuela-Toledo}}]{Almeida:2019hhx}%
  \BibitemOpen
  \bibfield  {author} {\bibinfo {author} {\bibfnamefont {J.~P.}\ \bibnamefont
  {Beltr\'an~Almeida}}, \bibinfo {author} {\bibfnamefont {J.}~\bibnamefont
  {Motoa-Manzano}},\ and\ \bibinfo {author} {\bibfnamefont {C.~A.}\
  \bibnamefont {Valenzuela-Toledo}},\ }\href
  {https://doi.org/10.1007/JHEP09(2019)118} {\bibfield  {journal} {\bibinfo
  {journal} {JHEP}\ }\textbf {\bibinfo {volume} {09}},\ \bibinfo {pages}
  {118}},\ \Eprint {https://arxiv.org/abs/1905.00900} {arXiv:1905.00900
  [gr-qc]} \BibitemShut {NoStop}%
\bibitem [{\citenamefont {Domcke}\ \emph
  {et~al.}(2020{\natexlab{a}})\citenamefont {Domcke}, \citenamefont {Ema},\
  and\ \citenamefont {Mukaida}}]{Domcke:2019qmm}%
  \BibitemOpen
  \bibfield  {author} {\bibinfo {author} {\bibfnamefont {V.}~\bibnamefont
  {Domcke}}, \bibinfo {author} {\bibfnamefont {Y.}~\bibnamefont {Ema}},\ and\
  \bibinfo {author} {\bibfnamefont {K.}~\bibnamefont {Mukaida}},\ }\href
  {https://doi.org/10.1007/JHEP02(2020)055} {\bibfield  {journal} {\bibinfo
  {journal} {JHEP}\ }\textbf {\bibinfo {volume} {02}},\ \bibinfo {pages}
  {055}},\ \Eprint {https://arxiv.org/abs/1910.01205} {arXiv:1910.01205
  [hep-ph]} \BibitemShut {NoStop}%
\bibitem [{\citenamefont {Domcke}\ \emph
  {et~al.}(2020{\natexlab{b}})\citenamefont {Domcke}, \citenamefont {Guidetti},
  \citenamefont {Welling},\ and\ \citenamefont {Westphal}}]{Domcke:2020zez}%
  \BibitemOpen
  \bibfield  {author} {\bibinfo {author} {\bibfnamefont {V.}~\bibnamefont
  {Domcke}}, \bibinfo {author} {\bibfnamefont {V.}~\bibnamefont {Guidetti}},
  \bibinfo {author} {\bibfnamefont {Y.}~\bibnamefont {Welling}},\ and\ \bibinfo
  {author} {\bibfnamefont {A.}~\bibnamefont {Westphal}},\ }\href
  {https://doi.org/10.1088/1475-7516/2020/09/009} {\bibfield  {journal}
  {\bibinfo  {journal} {JCAP}\ }\textbf {\bibinfo {volume} {09}},\ \bibinfo
  {pages} {009}},\ \Eprint {https://arxiv.org/abs/2002.02952} {arXiv:2002.02952
  [astro-ph.CO]} \BibitemShut {NoStop}%
\bibitem [{\citenamefont {Durrer}\ \emph {et~al.}(2011)\citenamefont {Durrer},
  \citenamefont {Hollenstein},\ and\ \citenamefont {Jain}}]{Durrer:2010mq}%
  \BibitemOpen
  \bibfield  {author} {\bibinfo {author} {\bibfnamefont {R.}~\bibnamefont
  {Durrer}}, \bibinfo {author} {\bibfnamefont {L.}~\bibnamefont
  {Hollenstein}},\ and\ \bibinfo {author} {\bibfnamefont {R.~K.}\ \bibnamefont
  {Jain}},\ }\href {https://doi.org/10.1088/1475-7516/2011/03/037} {\bibfield
  {journal} {\bibinfo  {journal} {JCAP}\ }\textbf {\bibinfo {volume} {03}},\
  \bibinfo {pages} {037}},\ \Eprint {https://arxiv.org/abs/1005.5322}
  {arXiv:1005.5322 [astro-ph.CO]} \BibitemShut {NoStop}%
\bibitem [{\citenamefont {Ng}\ \emph {et~al.}(2015)\citenamefont {Ng},
  \citenamefont {Cheng},\ and\ \citenamefont {Lee}}]{Ng:2015ewp}%
  \BibitemOpen
  \bibfield  {author} {\bibinfo {author} {\bibfnamefont {K.-W.}\ \bibnamefont
  {Ng}}, \bibinfo {author} {\bibfnamefont {S.-L.}\ \bibnamefont {Cheng}},\ and\
  \bibinfo {author} {\bibfnamefont {W.}~\bibnamefont {Lee}},\ }\href
  {https://doi.org/10.6122/CJP.20150909} {\bibfield  {journal} {\bibinfo
  {journal} {Chin. J. Phys.}\ }\textbf {\bibinfo {volume} {53}},\ \bibinfo
  {pages} {110105} (\bibinfo {year} {2015})},\ \Eprint
  {https://arxiv.org/abs/1409.2656} {arXiv:1409.2656 [astro-ph.CO]}
  \BibitemShut {NoStop}%
\bibitem [{\citenamefont {Fujita}\ \emph {et~al.}(2015)\citenamefont {Fujita},
  \citenamefont {Namba}, \citenamefont {Tada}, \citenamefont {Takeda},\ and\
  \citenamefont {Tashiro}}]{Fujita:2015iga}%
  \BibitemOpen
  \bibfield  {author} {\bibinfo {author} {\bibfnamefont {T.}~\bibnamefont
  {Fujita}}, \bibinfo {author} {\bibfnamefont {R.}~\bibnamefont {Namba}},
  \bibinfo {author} {\bibfnamefont {Y.}~\bibnamefont {Tada}}, \bibinfo {author}
  {\bibfnamefont {N.}~\bibnamefont {Takeda}},\ and\ \bibinfo {author}
  {\bibfnamefont {H.}~\bibnamefont {Tashiro}},\ }\href
  {https://doi.org/10.1088/1475-7516/2015/05/054} {\bibfield  {journal}
  {\bibinfo  {journal} {JCAP}\ }\textbf {\bibinfo {volume} {05}},\ \bibinfo
  {pages} {054}},\ \Eprint {https://arxiv.org/abs/1503.05802} {arXiv:1503.05802
  [astro-ph.CO]} \BibitemShut {NoStop}%
\bibitem [{\citenamefont {Adshead}\ \emph {et~al.}(2016)\citenamefont
  {Adshead}, \citenamefont {Giblin}, \citenamefont {Scully},\ and\
  \citenamefont {Sfakianakis}}]{Adshead:2016iae}%
  \BibitemOpen
  \bibfield  {author} {\bibinfo {author} {\bibfnamefont {P.}~\bibnamefont
  {Adshead}}, \bibinfo {author} {\bibfnamefont {J.~T.}\ \bibnamefont {Giblin}},
  \bibinfo {author} {\bibfnamefont {T.~R.}\ \bibnamefont {Scully}},\ and\
  \bibinfo {author} {\bibfnamefont {E.~I.}\ \bibnamefont {Sfakianakis}},\
  }\href {https://doi.org/10.1088/1475-7516/2016/10/039} {\bibfield  {journal}
  {\bibinfo  {journal} {JCAP}\ }\textbf {\bibinfo {volume} {10}},\ \bibinfo
  {pages} {039}},\ \Eprint {https://arxiv.org/abs/1606.08474} {arXiv:1606.08474
  [astro-ph.CO]} \BibitemShut {NoStop}%
\bibitem [{\citenamefont {Caprini}\ \emph {et~al.}(2018)\citenamefont
  {Caprini}, \citenamefont {Guzzetti},\ and\ \citenamefont
  {Sorbo}}]{Caprini:2017vnn}%
  \BibitemOpen
  \bibfield  {author} {\bibinfo {author} {\bibfnamefont {C.}~\bibnamefont
  {Caprini}}, \bibinfo {author} {\bibfnamefont {M.~C.}\ \bibnamefont
  {Guzzetti}},\ and\ \bibinfo {author} {\bibfnamefont {L.}~\bibnamefont
  {Sorbo}},\ }\href {https://doi.org/10.1088/1361-6382/aac143} {\bibfield
  {journal} {\bibinfo  {journal} {Class. Quant. Grav.}\ }\textbf {\bibinfo
  {volume} {35}},\ \bibinfo {pages} {124003} (\bibinfo {year} {2018})},\
  \Eprint {https://arxiv.org/abs/1707.09750} {arXiv:1707.09750 [astro-ph.CO]}
  \BibitemShut {NoStop}%
\bibitem [{\citenamefont {Shtanov}(2019)}]{Shtanov:2019civ}%
  \BibitemOpen
  \bibfield  {author} {\bibinfo {author} {\bibfnamefont {Y.}~\bibnamefont
  {Shtanov}},\ }\href {https://doi.org/10.1088/1475-7516/2019/10/008}
  {\bibfield  {journal} {\bibinfo  {journal} {JCAP}\ }\textbf {\bibinfo
  {volume} {10}},\ \bibinfo {pages} {008}},\ \Eprint
  {https://arxiv.org/abs/1902.05894} {arXiv:1902.05894 [astro-ph.CO]}
  \BibitemShut {NoStop}%
\bibitem [{\citenamefont {Shtanov}\ and\ \citenamefont
  {Pavliuk}(2019)}]{Shtanov:2019gpx}%
  \BibitemOpen
  \bibfield  {author} {\bibinfo {author} {\bibfnamefont {Y.~V.}\ \bibnamefont
  {Shtanov}}\ and\ \bibinfo {author} {\bibfnamefont {M.~V.}\ \bibnamefont
  {Pavliuk}},\ }\href {https://doi.org/10.15407/ujpe64.11.1009} {\bibfield
  {journal} {\bibinfo  {journal} {Ukr. Phys. J.}\ }\textbf {\bibinfo {volume}
  {64}},\ \bibinfo {pages} {1009} (\bibinfo {year} {2019})},\ \Eprint
  {https://arxiv.org/abs/1911.10424} {arXiv:1911.10424 [astro-ph.CO]}
  \BibitemShut {NoStop}%
\bibitem [{\citenamefont {Patel}\ \emph {et~al.}(2020)\citenamefont {Patel},
  \citenamefont {Tashiro},\ and\ \citenamefont {Urakawa}}]{Patel:2019isj}%
  \BibitemOpen
  \bibfield  {author} {\bibinfo {author} {\bibfnamefont {T.}~\bibnamefont
  {Patel}}, \bibinfo {author} {\bibfnamefont {H.}~\bibnamefont {Tashiro}},\
  and\ \bibinfo {author} {\bibfnamefont {Y.}~\bibnamefont {Urakawa}},\ }\href
  {https://doi.org/10.1088/1475-7516/2020/01/043} {\bibfield  {journal}
  {\bibinfo  {journal} {JCAP}\ }\textbf {\bibinfo {volume} {01}},\ \bibinfo
  {pages} {043}},\ \Eprint {https://arxiv.org/abs/1909.00288} {arXiv:1909.00288
  [astro-ph.CO]} \BibitemShut {NoStop}%
\bibitem [{\citenamefont {Fujita}\ and\ \citenamefont
  {Durrer}(2019)}]{Fujita:2019pmi}%
  \BibitemOpen
  \bibfield  {author} {\bibinfo {author} {\bibfnamefont {T.}~\bibnamefont
  {Fujita}}\ and\ \bibinfo {author} {\bibfnamefont {R.}~\bibnamefont
  {Durrer}},\ }\href {https://doi.org/10.1088/1475-7516/2019/09/008} {\bibfield
   {journal} {\bibinfo  {journal} {JCAP}\ }\textbf {\bibinfo {volume} {09}},\
  \bibinfo {pages} {008}},\ \Eprint {https://arxiv.org/abs/1904.11428}
  {arXiv:1904.11428 [astro-ph.CO]} \BibitemShut {NoStop}%
\bibitem [{\citenamefont {Sobol}\ \emph {et~al.}(2019)\citenamefont {Sobol},
  \citenamefont {Gorbar},\ and\ \citenamefont {Vilchinskii}}]{Sobol:2019xls}%
  \BibitemOpen
  \bibfield  {author} {\bibinfo {author} {\bibfnamefont {O.~O.}\ \bibnamefont
  {Sobol}}, \bibinfo {author} {\bibfnamefont {E.~V.}\ \bibnamefont {Gorbar}},\
  and\ \bibinfo {author} {\bibfnamefont {S.~I.}\ \bibnamefont {Vilchinskii}},\
  }\href {https://doi.org/10.1103/PhysRevD.100.063523} {\bibfield  {journal}
  {\bibinfo  {journal} {Phys. Rev. D}\ }\textbf {\bibinfo {volume} {100}},\
  \bibinfo {pages} {063523} (\bibinfo {year} {2019})},\ \Eprint
  {https://arxiv.org/abs/1907.10443} {arXiv:1907.10443 [astro-ph.CO]}
  \BibitemShut {NoStop}%
\bibitem [{\citenamefont {Jim\'enez}\ \emph {et~al.}(2017)\citenamefont
  {Jim\'enez}, \citenamefont {Kamada}, \citenamefont {Schmitz},\ and\
  \citenamefont {Xu}}]{Jimenez:2017cdr}%
  \BibitemOpen
  \bibfield  {author} {\bibinfo {author} {\bibfnamefont {D.}~\bibnamefont
  {Jim\'enez}}, \bibinfo {author} {\bibfnamefont {K.}~\bibnamefont {Kamada}},
  \bibinfo {author} {\bibfnamefont {K.}~\bibnamefont {Schmitz}},\ and\ \bibinfo
  {author} {\bibfnamefont {X.-J.}\ \bibnamefont {Xu}},\ }\href
  {https://doi.org/10.1088/1475-7516/2017/12/011} {\bibfield  {journal}
  {\bibinfo  {journal} {JCAP}\ }\textbf {\bibinfo {volume} {12}},\ \bibinfo
  {pages} {011}},\ \Eprint {https://arxiv.org/abs/1707.07943} {arXiv:1707.07943
  [hep-ph]} \BibitemShut {NoStop}%
\bibitem [{\citenamefont {Bugaev}\ and\ \citenamefont
  {Klimai}(2014)}]{Bugaev:2013fya}%
  \BibitemOpen
  \bibfield  {author} {\bibinfo {author} {\bibfnamefont {E.}~\bibnamefont
  {Bugaev}}\ and\ \bibinfo {author} {\bibfnamefont {P.}~\bibnamefont
  {Klimai}},\ }\href {https://doi.org/10.1103/PhysRevD.90.103501} {\bibfield
  {journal} {\bibinfo  {journal} {Phys. Rev. D}\ }\textbf {\bibinfo {volume}
  {90}},\ \bibinfo {pages} {103501} (\bibinfo {year} {2014})},\ \Eprint
  {https://arxiv.org/abs/1312.7435} {arXiv:1312.7435 [astro-ph.CO]}
  \BibitemShut {NoStop}%
\bibitem [{\citenamefont {Erfani}(2016)}]{Erfani:2015rqv}%
  \BibitemOpen
  \bibfield  {author} {\bibinfo {author} {\bibfnamefont {E.}~\bibnamefont
  {Erfani}},\ }\href {https://doi.org/10.1088/1475-7516/2016/04/020} {\bibfield
   {journal} {\bibinfo  {journal} {JCAP}\ }\textbf {\bibinfo {volume} {04}},\
  \bibinfo {pages} {020}},\ \Eprint {https://arxiv.org/abs/1511.08470}
  {arXiv:1511.08470 [astro-ph.CO]} \BibitemShut {NoStop}%
\bibitem [{\citenamefont {Domcke}\ \emph {et~al.}(2017)\citenamefont {Domcke},
  \citenamefont {Muia}, \citenamefont {Pieroni},\ and\ \citenamefont
  {Witkowski}}]{Domcke:2017fix}%
  \BibitemOpen
  \bibfield  {author} {\bibinfo {author} {\bibfnamefont {V.}~\bibnamefont
  {Domcke}}, \bibinfo {author} {\bibfnamefont {F.}~\bibnamefont {Muia}},
  \bibinfo {author} {\bibfnamefont {M.}~\bibnamefont {Pieroni}},\ and\ \bibinfo
  {author} {\bibfnamefont {L.~T.}\ \bibnamefont {Witkowski}},\ }\href
  {https://doi.org/10.1088/1475-7516/2017/07/048} {\bibfield  {journal}
  {\bibinfo  {journal} {JCAP}\ }\textbf {\bibinfo {volume} {07}},\ \bibinfo
  {pages} {048}},\ \Eprint {https://arxiv.org/abs/1704.03464} {arXiv:1704.03464
  [astro-ph.CO]} \BibitemShut {NoStop}%
\bibitem [{\citenamefont {Cheng}\ \emph {et~al.}(2018)\citenamefont {Cheng},
  \citenamefont {Lee},\ and\ \citenamefont {Ng}}]{Cheng:2018yyr}%
  \BibitemOpen
  \bibfield  {author} {\bibinfo {author} {\bibfnamefont {S.-L.}\ \bibnamefont
  {Cheng}}, \bibinfo {author} {\bibfnamefont {W.}~\bibnamefont {Lee}},\ and\
  \bibinfo {author} {\bibfnamefont {K.-W.}\ \bibnamefont {Ng}},\ }\href
  {https://doi.org/10.1088/1475-7516/2018/07/001} {\bibfield  {journal}
  {\bibinfo  {journal} {JCAP}\ }\textbf {\bibinfo {volume} {07}},\ \bibinfo
  {pages} {001}},\ \Eprint {https://arxiv.org/abs/1801.09050} {arXiv:1801.09050
  [astro-ph.CO]} \BibitemShut {NoStop}%
\bibitem [{\citenamefont {\"Ozsoy}\ and\ \citenamefont
  {Lalak}(2021)}]{Ozsoy:2020kat}%
  \BibitemOpen
  \bibfield  {author} {\bibinfo {author} {\bibfnamefont {O.}~\bibnamefont
  {\"Ozsoy}}\ and\ \bibinfo {author} {\bibfnamefont {Z.}~\bibnamefont
  {Lalak}},\ }\href {https://doi.org/10.1088/1475-7516/2021/01/040} {\bibfield
  {journal} {\bibinfo  {journal} {JCAP}\ }\textbf {\bibinfo {volume} {01}},\
  \bibinfo {pages} {040}},\ \Eprint {https://arxiv.org/abs/2008.07549}
  {arXiv:2008.07549 [astro-ph.CO]} \BibitemShut {NoStop}%
\bibitem [{\citenamefont {Kamada}\ and\ \citenamefont
  {Nakai}(2017)}]{Kamada:2017cpk}%
  \BibitemOpen
  \bibfield  {author} {\bibinfo {author} {\bibfnamefont {K.}~\bibnamefont
  {Kamada}}\ and\ \bibinfo {author} {\bibfnamefont {Y.}~\bibnamefont {Nakai}},\
  }\href {https://doi.org/10.1103/PhysRevD.96.023537} {\bibfield  {journal}
  {\bibinfo  {journal} {Phys. Rev. D}\ }\textbf {\bibinfo {volume} {96}},\
  \bibinfo {pages} {023537} (\bibinfo {year} {2017})},\ \Eprint
  {https://arxiv.org/abs/1702.03928} {arXiv:1702.03928 [hep-ph]} \BibitemShut
  {NoStop}%
\bibitem [{\citenamefont {Agrawal}\ \emph
  {et~al.}(2018{\natexlab{a}})\citenamefont {Agrawal}, \citenamefont
  {Marques-Tavares},\ and\ \citenamefont {Xue}}]{Agrawal:2017eqm}%
  \BibitemOpen
  \bibfield  {author} {\bibinfo {author} {\bibfnamefont {P.}~\bibnamefont
  {Agrawal}}, \bibinfo {author} {\bibfnamefont {G.}~\bibnamefont
  {Marques-Tavares}},\ and\ \bibinfo {author} {\bibfnamefont {W.}~\bibnamefont
  {Xue}},\ }\href {https://doi.org/10.1007/JHEP03(2018)049} {\bibfield
  {journal} {\bibinfo  {journal} {JHEP}\ }\textbf {\bibinfo {volume} {03}},\
  \bibinfo {pages} {049}},\ \Eprint {https://arxiv.org/abs/1708.05008}
  {arXiv:1708.05008 [hep-ph]} \BibitemShut {NoStop}%
\bibitem [{\citenamefont {Co}\ \emph {et~al.}(2019)\citenamefont {Co},
  \citenamefont {Pierce}, \citenamefont {Zhang},\ and\ \citenamefont
  {Zhao}}]{Co:2018lka}%
  \BibitemOpen
  \bibfield  {author} {\bibinfo {author} {\bibfnamefont {R.~T.}\ \bibnamefont
  {Co}}, \bibinfo {author} {\bibfnamefont {A.}~\bibnamefont {Pierce}}, \bibinfo
  {author} {\bibfnamefont {Z.}~\bibnamefont {Zhang}},\ and\ \bibinfo {author}
  {\bibfnamefont {Y.}~\bibnamefont {Zhao}},\ }\href
  {https://doi.org/10.1103/PhysRevD.99.075002} {\bibfield  {journal} {\bibinfo
  {journal} {Phys. Rev. D}\ }\textbf {\bibinfo {volume} {99}},\ \bibinfo
  {pages} {075002} (\bibinfo {year} {2019})},\ \Eprint
  {https://arxiv.org/abs/1810.07196} {arXiv:1810.07196 [hep-ph]} \BibitemShut
  {NoStop}%
\bibitem [{\citenamefont {Bastero-Gil}\ \emph {et~al.}(2019)\citenamefont
  {Bastero-Gil}, \citenamefont {Santiago}, \citenamefont {Ubaldi},\ and\
  \citenamefont {Vega-Morales}}]{Bastero-Gil:2018uel}%
  \BibitemOpen
  \bibfield  {author} {\bibinfo {author} {\bibfnamefont {M.}~\bibnamefont
  {Bastero-Gil}}, \bibinfo {author} {\bibfnamefont {J.}~\bibnamefont
  {Santiago}}, \bibinfo {author} {\bibfnamefont {L.}~\bibnamefont {Ubaldi}},\
  and\ \bibinfo {author} {\bibfnamefont {R.}~\bibnamefont {Vega-Morales}},\
  }\href {https://doi.org/10.1088/1475-7516/2019/04/015} {\bibfield  {journal}
  {\bibinfo  {journal} {JCAP}\ }\textbf {\bibinfo {volume} {04}},\ \bibinfo
  {pages} {015}},\ \Eprint {https://arxiv.org/abs/1810.07208} {arXiv:1810.07208
  [hep-ph]} \BibitemShut {NoStop}%
\bibitem [{\citenamefont {Agrawal}\ \emph {et~al.}(2020)\citenamefont
  {Agrawal}, \citenamefont {Kitajima}, \citenamefont {Reece}, \citenamefont
  {Sekiguchi},\ and\ \citenamefont {Takahashi}}]{Agrawal:2018vin}%
  \BibitemOpen
  \bibfield  {author} {\bibinfo {author} {\bibfnamefont {P.}~\bibnamefont
  {Agrawal}}, \bibinfo {author} {\bibfnamefont {N.}~\bibnamefont {Kitajima}},
  \bibinfo {author} {\bibfnamefont {M.}~\bibnamefont {Reece}}, \bibinfo
  {author} {\bibfnamefont {T.}~\bibnamefont {Sekiguchi}},\ and\ \bibinfo
  {author} {\bibfnamefont {F.}~\bibnamefont {Takahashi}},\ }\href
  {https://doi.org/10.1016/j.physletb.2019.135136} {\bibfield  {journal}
  {\bibinfo  {journal} {Phys. Lett. B}\ }\textbf {\bibinfo {volume} {801}},\
  \bibinfo {pages} {135136} (\bibinfo {year} {2020})},\ \Eprint
  {https://arxiv.org/abs/1810.07188} {arXiv:1810.07188 [hep-ph]} \BibitemShut
  {NoStop}%
\bibitem [{\citenamefont {Machado}\ \emph {et~al.}(2019)\citenamefont
  {Machado}, \citenamefont {Ratzinger}, \citenamefont {Schwaller},\ and\
  \citenamefont {Stefanek}}]{Machado:2018nqk}%
  \BibitemOpen
  \bibfield  {author} {\bibinfo {author} {\bibfnamefont {C.~S.}\ \bibnamefont
  {Machado}}, \bibinfo {author} {\bibfnamefont {W.}~\bibnamefont {Ratzinger}},
  \bibinfo {author} {\bibfnamefont {P.}~\bibnamefont {Schwaller}},\ and\
  \bibinfo {author} {\bibfnamefont {B.~A.}\ \bibnamefont {Stefanek}},\ }\href
  {https://doi.org/10.1007/JHEP01(2019)053} {\bibfield  {journal} {\bibinfo
  {journal} {JHEP}\ }\textbf {\bibinfo {volume} {01}},\ \bibinfo {pages}
  {053}},\ \Eprint {https://arxiv.org/abs/1811.01950} {arXiv:1811.01950
  [hep-ph]} \BibitemShut {NoStop}%
\bibitem [{\citenamefont {Barnaby}\ \emph
  {et~al.}(2012{\natexlab{a}})\citenamefont {Barnaby}, \citenamefont {Moxon},
  \citenamefont {Namba}, \citenamefont {Peloso}, \citenamefont {Shiu},\ and\
  \citenamefont {Zhou}}]{Barnaby:2012xt}%
  \BibitemOpen
  \bibfield  {author} {\bibinfo {author} {\bibfnamefont {N.}~\bibnamefont
  {Barnaby}}, \bibinfo {author} {\bibfnamefont {J.}~\bibnamefont {Moxon}},
  \bibinfo {author} {\bibfnamefont {R.}~\bibnamefont {Namba}}, \bibinfo
  {author} {\bibfnamefont {M.}~\bibnamefont {Peloso}}, \bibinfo {author}
  {\bibfnamefont {G.}~\bibnamefont {Shiu}},\ and\ \bibinfo {author}
  {\bibfnamefont {P.}~\bibnamefont {Zhou}},\ }\href
  {https://doi.org/10.1103/PhysRevD.86.103508} {\bibfield  {journal} {\bibinfo
  {journal} {Phys. Rev. D}\ }\textbf {\bibinfo {volume} {86}},\ \bibinfo
  {pages} {103508} (\bibinfo {year} {2012}{\natexlab{a}})},\ \Eprint
  {https://arxiv.org/abs/1206.6117} {arXiv:1206.6117 [astro-ph.CO]}
  \BibitemShut {NoStop}%
\bibitem [{\citenamefont {Cook}\ and\ \citenamefont
  {Sorbo}(2013)}]{Cook:2013xea}%
  \BibitemOpen
  \bibfield  {author} {\bibinfo {author} {\bibfnamefont {J.~L.}\ \bibnamefont
  {Cook}}\ and\ \bibinfo {author} {\bibfnamefont {L.}~\bibnamefont {Sorbo}},\
  }\href {https://doi.org/10.1088/1475-7516/2013/11/047} {\bibfield  {journal}
  {\bibinfo  {journal} {JCAP}\ }\textbf {\bibinfo {volume} {11}},\ \bibinfo
  {pages} {047}},\ \Eprint {https://arxiv.org/abs/1307.7077} {arXiv:1307.7077
  [astro-ph.CO]} \BibitemShut {NoStop}%
\bibitem [{\citenamefont {Shiraishi}\ \emph {et~al.}(2013)\citenamefont
  {Shiraishi}, \citenamefont {Ricciardone},\ and\ \citenamefont
  {Saga}}]{Shiraishi:2013kxa}%
  \BibitemOpen
  \bibfield  {author} {\bibinfo {author} {\bibfnamefont {M.}~\bibnamefont
  {Shiraishi}}, \bibinfo {author} {\bibfnamefont {A.}~\bibnamefont
  {Ricciardone}},\ and\ \bibinfo {author} {\bibfnamefont {S.}~\bibnamefont
  {Saga}},\ }\href {https://doi.org/10.1088/1475-7516/2013/11/051} {\bibfield
  {journal} {\bibinfo  {journal} {JCAP}\ }\textbf {\bibinfo {volume} {11}},\
  \bibinfo {pages} {051}},\ \Eprint {https://arxiv.org/abs/1308.6769}
  {arXiv:1308.6769 [astro-ph.CO]} \BibitemShut {NoStop}%
\bibitem [{\citenamefont {Mukohyama}\ \emph {et~al.}(2014)\citenamefont
  {Mukohyama}, \citenamefont {Namba}, \citenamefont {Peloso},\ and\
  \citenamefont {Shiu}}]{Mukohyama:2014gba}%
  \BibitemOpen
  \bibfield  {author} {\bibinfo {author} {\bibfnamefont {S.}~\bibnamefont
  {Mukohyama}}, \bibinfo {author} {\bibfnamefont {R.}~\bibnamefont {Namba}},
  \bibinfo {author} {\bibfnamefont {M.}~\bibnamefont {Peloso}},\ and\ \bibinfo
  {author} {\bibfnamefont {G.}~\bibnamefont {Shiu}},\ }\href
  {https://doi.org/10.1088/1475-7516/2014/08/036} {\bibfield  {journal}
  {\bibinfo  {journal} {JCAP}\ }\textbf {\bibinfo {volume} {08}},\ \bibinfo
  {pages} {036}},\ \Eprint {https://arxiv.org/abs/1405.0346} {arXiv:1405.0346
  [astro-ph.CO]} \BibitemShut {NoStop}%
\bibitem [{\citenamefont {Mirbabayi}\ \emph {et~al.}(2015)\citenamefont
  {Mirbabayi}, \citenamefont {Senatore}, \citenamefont {Silverstein},\ and\
  \citenamefont {Zaldarriaga}}]{Mirbabayi:2014jqa}%
  \BibitemOpen
  \bibfield  {author} {\bibinfo {author} {\bibfnamefont {M.}~\bibnamefont
  {Mirbabayi}}, \bibinfo {author} {\bibfnamefont {L.}~\bibnamefont {Senatore}},
  \bibinfo {author} {\bibfnamefont {E.}~\bibnamefont {Silverstein}},\ and\
  \bibinfo {author} {\bibfnamefont {M.}~\bibnamefont {Zaldarriaga}},\ }\href
  {https://doi.org/10.1103/PhysRevD.91.063518} {\bibfield  {journal} {\bibinfo
  {journal} {Phys. Rev. D}\ }\textbf {\bibinfo {volume} {91}},\ \bibinfo
  {pages} {063518} (\bibinfo {year} {2015})},\ \Eprint
  {https://arxiv.org/abs/1412.0665} {arXiv:1412.0665 [hep-th]} \BibitemShut
  {NoStop}%
\bibitem [{\citenamefont {Namba}\ \emph {et~al.}(2016)\citenamefont {Namba},
  \citenamefont {Peloso}, \citenamefont {Shiraishi}, \citenamefont {Sorbo},\
  and\ \citenamefont {Unal}}]{Namba:2015gja}%
  \BibitemOpen
  \bibfield  {author} {\bibinfo {author} {\bibfnamefont {R.}~\bibnamefont
  {Namba}}, \bibinfo {author} {\bibfnamefont {M.}~\bibnamefont {Peloso}},
  \bibinfo {author} {\bibfnamefont {M.}~\bibnamefont {Shiraishi}}, \bibinfo
  {author} {\bibfnamefont {L.}~\bibnamefont {Sorbo}},\ and\ \bibinfo {author}
  {\bibfnamefont {C.}~\bibnamefont {Unal}},\ }\href
  {https://doi.org/10.1088/1475-7516/2016/01/041} {\bibfield  {journal}
  {\bibinfo  {journal} {JCAP}\ }\textbf {\bibinfo {volume} {01}},\ \bibinfo
  {pages} {041}},\ \Eprint {https://arxiv.org/abs/1509.07521} {arXiv:1509.07521
  [astro-ph.CO]} \BibitemShut {NoStop}%
\bibitem [{\citenamefont {Domcke}\ \emph {et~al.}(2016)\citenamefont {Domcke},
  \citenamefont {Pieroni},\ and\ \citenamefont {Bin\'etruy}}]{Domcke:2016bkh}%
  \BibitemOpen
  \bibfield  {author} {\bibinfo {author} {\bibfnamefont {V.}~\bibnamefont
  {Domcke}}, \bibinfo {author} {\bibfnamefont {M.}~\bibnamefont {Pieroni}},\
  and\ \bibinfo {author} {\bibfnamefont {P.}~\bibnamefont {Bin\'etruy}},\
  }\href {https://doi.org/10.1088/1475-7516/2016/06/031} {\bibfield  {journal}
  {\bibinfo  {journal} {JCAP}\ }\textbf {\bibinfo {volume} {06}},\ \bibinfo
  {pages} {031}},\ \Eprint {https://arxiv.org/abs/1603.01287} {arXiv:1603.01287
  [astro-ph.CO]} \BibitemShut {NoStop}%
\bibitem [{\citenamefont {Shiraishi}\ \emph {et~al.}(2016)\citenamefont
  {Shiraishi}, \citenamefont {Hikage}, \citenamefont {Namba}, \citenamefont
  {Namikawa},\ and\ \citenamefont {Hazumi}}]{Shiraishi:2016yun}%
  \BibitemOpen
  \bibfield  {author} {\bibinfo {author} {\bibfnamefont {M.}~\bibnamefont
  {Shiraishi}}, \bibinfo {author} {\bibfnamefont {C.}~\bibnamefont {Hikage}},
  \bibinfo {author} {\bibfnamefont {R.}~\bibnamefont {Namba}}, \bibinfo
  {author} {\bibfnamefont {T.}~\bibnamefont {Namikawa}},\ and\ \bibinfo
  {author} {\bibfnamefont {M.}~\bibnamefont {Hazumi}},\ }\href
  {https://doi.org/10.1103/PhysRevD.94.043506} {\bibfield  {journal} {\bibinfo
  {journal} {Phys. Rev. D}\ }\textbf {\bibinfo {volume} {94}},\ \bibinfo
  {pages} {043506} (\bibinfo {year} {2016})},\ \Eprint
  {https://arxiv.org/abs/1606.06082} {arXiv:1606.06082 [astro-ph.CO]}
  \BibitemShut {NoStop}%
\bibitem [{\citenamefont {Obata}(2017)}]{Obata:2016oym}%
  \BibitemOpen
  \bibfield  {author} {\bibinfo {author} {\bibfnamefont {I.}~\bibnamefont
  {Obata}},\ }\href {https://doi.org/10.1088/1475-7516/2017/06/050} {\bibfield
  {journal} {\bibinfo  {journal} {JCAP}\ }\textbf {\bibinfo {volume} {06}},\
  \bibinfo {pages} {050}},\ \Eprint {https://arxiv.org/abs/1612.08817}
  {arXiv:1612.08817 [astro-ph.CO]} \BibitemShut {NoStop}%
\bibitem [{\citenamefont {\"Ozsoy}(2021)}]{Ozsoy:2020ccy}%
  \BibitemOpen
  \bibfield  {author} {\bibinfo {author} {\bibfnamefont {O.}~\bibnamefont
  {\"Ozsoy}},\ }\href {https://doi.org/10.1088/1475-7516/2021/04/040}
  {\bibfield  {journal} {\bibinfo  {journal} {JCAP}\ }\textbf {\bibinfo
  {volume} {04}},\ \bibinfo {pages} {040}},\ \Eprint
  {https://arxiv.org/abs/2005.10280} {arXiv:2005.10280 [astro-ph.CO]}
  \BibitemShut {NoStop}%
\bibitem [{\citenamefont {Cook}\ and\ \citenamefont
  {Sorbo}(2012)}]{Cook:2011hg}%
  \BibitemOpen
  \bibfield  {author} {\bibinfo {author} {\bibfnamefont {J.~L.}\ \bibnamefont
  {Cook}}\ and\ \bibinfo {author} {\bibfnamefont {L.}~\bibnamefont {Sorbo}},\
  }\href {https://doi.org/10.1103/PhysRevD.85.023534} {\bibfield  {journal}
  {\bibinfo  {journal} {Phys. Rev. D}\ }\textbf {\bibinfo {volume} {85}},\
  \bibinfo {pages} {023534} (\bibinfo {year} {2012})},\ \bibinfo {note}
  {[Erratum: Phys.Rev.D 86, 069901 (2012)]},\ \Eprint
  {https://arxiv.org/abs/1109.0022} {arXiv:1109.0022 [astro-ph.CO]}
  \BibitemShut {NoStop}%
\bibitem [{\citenamefont {Barnaby}\ \emph
  {et~al.}(2012{\natexlab{b}})\citenamefont {Barnaby}, \citenamefont {Pajer},\
  and\ \citenamefont {Peloso}}]{Barnaby:2011qe}%
  \BibitemOpen
  \bibfield  {author} {\bibinfo {author} {\bibfnamefont {N.}~\bibnamefont
  {Barnaby}}, \bibinfo {author} {\bibfnamefont {E.}~\bibnamefont {Pajer}},\
  and\ \bibinfo {author} {\bibfnamefont {M.}~\bibnamefont {Peloso}},\ }\href
  {https://doi.org/10.1103/PhysRevD.85.023525} {\bibfield  {journal} {\bibinfo
  {journal} {Phys. Rev. D}\ }\textbf {\bibinfo {volume} {85}},\ \bibinfo
  {pages} {023525} (\bibinfo {year} {2012}{\natexlab{b}})},\ \Eprint
  {https://arxiv.org/abs/1110.3327} {arXiv:1110.3327 [astro-ph.CO]}
  \BibitemShut {NoStop}%
\bibitem [{\citenamefont {Crowder}\ \emph {et~al.}(2013)\citenamefont
  {Crowder}, \citenamefont {Namba}, \citenamefont {Mandic}, \citenamefont
  {Mukohyama},\ and\ \citenamefont {Peloso}}]{Crowder:2012ik}%
  \BibitemOpen
  \bibfield  {author} {\bibinfo {author} {\bibfnamefont {S.~G.}\ \bibnamefont
  {Crowder}}, \bibinfo {author} {\bibfnamefont {R.}~\bibnamefont {Namba}},
  \bibinfo {author} {\bibfnamefont {V.}~\bibnamefont {Mandic}}, \bibinfo
  {author} {\bibfnamefont {S.}~\bibnamefont {Mukohyama}},\ and\ \bibinfo
  {author} {\bibfnamefont {M.}~\bibnamefont {Peloso}},\ }\href
  {https://doi.org/10.1016/j.physletb.2013.08.077} {\bibfield  {journal}
  {\bibinfo  {journal} {Phys. Lett. B}\ }\textbf {\bibinfo {volume} {726}},\
  \bibinfo {pages} {66} (\bibinfo {year} {2013})},\ \Eprint
  {https://arxiv.org/abs/1212.4165} {arXiv:1212.4165 [astro-ph.CO]}
  \BibitemShut {NoStop}%
\bibitem [{\citenamefont {Garcia-Bellido}\ \emph {et~al.}(2016)\citenamefont
  {Garcia-Bellido}, \citenamefont {Peloso},\ and\ \citenamefont
  {Unal}}]{Garcia-Bellido:2016dkw}%
  \BibitemOpen
  \bibfield  {author} {\bibinfo {author} {\bibfnamefont {J.}~\bibnamefont
  {Garcia-Bellido}}, \bibinfo {author} {\bibfnamefont {M.}~\bibnamefont
  {Peloso}},\ and\ \bibinfo {author} {\bibfnamefont {C.}~\bibnamefont {Unal}},\
  }\href {https://doi.org/10.1088/1475-7516/2016/12/031} {\bibfield  {journal}
  {\bibinfo  {journal} {JCAP}\ }\textbf {\bibinfo {volume} {12}},\ \bibinfo
  {pages} {031}},\ \Eprint {https://arxiv.org/abs/1610.03763} {arXiv:1610.03763
  [astro-ph.CO]} \BibitemShut {NoStop}%
\bibitem [{\citenamefont {Obata}\ and\ \citenamefont
  {Soda}(2016{\natexlab{a}})}]{Obata:2016tmo}%
  \BibitemOpen
  \bibfield  {author} {\bibinfo {author} {\bibfnamefont {I.}~\bibnamefont
  {Obata}}\ and\ \bibinfo {author} {\bibfnamefont {J.}~\bibnamefont {Soda}},\
  }\href {https://doi.org/10.1103/PhysRevD.93.123502} {\bibfield  {journal}
  {\bibinfo  {journal} {Phys. Rev. D}\ }\textbf {\bibinfo {volume} {93}},\
  \bibinfo {pages} {123502} (\bibinfo {year} {2016}{\natexlab{a}})},\ \bibinfo
  {note} {[Addendum: Phys.Rev.D 95, 109903 (2017)]},\ \Eprint
  {https://arxiv.org/abs/1602.06024} {arXiv:1602.06024 [hep-th]} \BibitemShut
  {NoStop}%
\bibitem [{\citenamefont {Obata}\ and\ \citenamefont
  {Soda}(2016{\natexlab{b}})}]{Obata:2016xcr}%
  \BibitemOpen
  \bibfield  {author} {\bibinfo {author} {\bibfnamefont {I.}~\bibnamefont
  {Obata}}\ and\ \bibinfo {author} {\bibfnamefont {J.}~\bibnamefont {Soda}},\
  }\href {https://doi.org/10.1103/PhysRevD.94.044062} {\bibfield  {journal}
  {\bibinfo  {journal} {Phys. Rev. D}\ }\textbf {\bibinfo {volume} {94}},\
  \bibinfo {pages} {044062} (\bibinfo {year} {2016}{\natexlab{b}})},\ \Eprint
  {https://arxiv.org/abs/1607.01847} {arXiv:1607.01847 [astro-ph.CO]}
  \BibitemShut {NoStop}%
\bibitem [{\citenamefont {Machado}\ \emph {et~al.}(2020)\citenamefont
  {Machado}, \citenamefont {Ratzinger}, \citenamefont {Schwaller},\ and\
  \citenamefont {Stefanek}}]{Machado:2019xuc}%
  \BibitemOpen
  \bibfield  {author} {\bibinfo {author} {\bibfnamefont {C.~S.}\ \bibnamefont
  {Machado}}, \bibinfo {author} {\bibfnamefont {W.}~\bibnamefont {Ratzinger}},
  \bibinfo {author} {\bibfnamefont {P.}~\bibnamefont {Schwaller}},\ and\
  \bibinfo {author} {\bibfnamefont {B.~A.}\ \bibnamefont {Stefanek}},\ }\href
  {https://doi.org/10.1103/PhysRevD.102.075033} {\bibfield  {journal} {\bibinfo
   {journal} {Phys. Rev. D}\ }\textbf {\bibinfo {volume} {102}},\ \bibinfo
  {pages} {075033} (\bibinfo {year} {2020})},\ \Eprint
  {https://arxiv.org/abs/1912.01007} {arXiv:1912.01007 [hep-ph]} \BibitemShut
  {NoStop}%
\bibitem [{\citenamefont {Okano}\ and\ \citenamefont
  {Fujita}(2021)}]{Okano:2020uyr}%
  \BibitemOpen
  \bibfield  {author} {\bibinfo {author} {\bibfnamefont {S.}~\bibnamefont
  {Okano}}\ and\ \bibinfo {author} {\bibfnamefont {T.}~\bibnamefont {Fujita}},\
  }\href {https://doi.org/10.1088/1475-7516/2021/03/026} {\bibfield  {journal}
  {\bibinfo  {journal} {JCAP}\ }\textbf {\bibinfo {volume} {03}},\ \bibinfo
  {pages} {026}},\ \Eprint {https://arxiv.org/abs/2005.13833} {arXiv:2005.13833
  [astro-ph.CO]} \BibitemShut {NoStop}%
\bibitem [{\citenamefont {Bartolo}\ \emph {et~al.}(2016)\citenamefont {Bartolo}
  \emph {et~al.}}]{Bartolo:2016ami}%
  \BibitemOpen
  \bibfield  {author} {\bibinfo {author} {\bibfnamefont {N.}~\bibnamefont
  {Bartolo}} \emph {et~al.},\ }\href
  {https://doi.org/10.1088/1475-7516/2016/12/026} {\bibfield  {journal}
  {\bibinfo  {journal} {JCAP}\ }\textbf {\bibinfo {volume} {12}},\ \bibinfo
  {pages} {026}},\ \Eprint {https://arxiv.org/abs/1610.06481} {arXiv:1610.06481
  [astro-ph.CO]} \BibitemShut {NoStop}%
\bibitem [{\citenamefont {Co}\ \emph {et~al.}(2021)\citenamefont {Co},
  \citenamefont {Harigaya},\ and\ \citenamefont {Pierce}}]{Co:2021rhi}%
  \BibitemOpen
  \bibfield  {author} {\bibinfo {author} {\bibfnamefont {R.~T.}\ \bibnamefont
  {Co}}, \bibinfo {author} {\bibfnamefont {K.}~\bibnamefont {Harigaya}},\ and\
  \bibinfo {author} {\bibfnamefont {A.}~\bibnamefont {Pierce}},\ }\href
  {https://doi.org/10.1007/JHEP12(2021)099} {\bibfield  {journal} {\bibinfo
  {journal} {JHEP}\ }\textbf {\bibinfo {volume} {12}},\ \bibinfo {pages}
  {099}},\ \Eprint {https://arxiv.org/abs/2104.02077} {arXiv:2104.02077
  [hep-ph]} \BibitemShut {NoStop}%
\bibitem [{\citenamefont {Ade}\ \emph {et~al.}(2014)\citenamefont {Ade} \emph
  {et~al.}}]{Ade:2013ydc}%
  \BibitemOpen
  \bibfield  {author} {\bibinfo {author} {\bibfnamefont {P.~A.~R.}\
  \bibnamefont {Ade}} \emph {et~al.} (\bibinfo {collaboration} {Planck}),\
  }\href {https://doi.org/10.1051/0004-6361/201321554} {\bibfield  {journal}
  {\bibinfo  {journal} {Astron. Astrophys.}\ }\textbf {\bibinfo {volume}
  {571}},\ \bibinfo {pages} {A24} (\bibinfo {year} {2014})},\ \Eprint
  {https://arxiv.org/abs/1303.5084} {arXiv:1303.5084 [astro-ph.CO]}
  \BibitemShut {NoStop}%
\bibitem [{\citenamefont {Ade}\ \emph {et~al.}(2016{\natexlab{a}})\citenamefont
  {Ade} \emph {et~al.}}]{Ade:2015lrj}%
  \BibitemOpen
  \bibfield  {author} {\bibinfo {author} {\bibfnamefont {P.~A.~R.}\
  \bibnamefont {Ade}} \emph {et~al.} (\bibinfo {collaboration} {Planck}),\
  }\href {https://doi.org/10.1051/0004-6361/201525898} {\bibfield  {journal}
  {\bibinfo  {journal} {Astron. Astrophys.}\ }\textbf {\bibinfo {volume}
  {594}},\ \bibinfo {pages} {A20} (\bibinfo {year} {2016}{\natexlab{a}})},\
  \Eprint {https://arxiv.org/abs/1502.02114} {arXiv:1502.02114 [astro-ph.CO]}
  \BibitemShut {NoStop}%
\bibitem [{\citenamefont {Ade}\ \emph {et~al.}(2016{\natexlab{b}})\citenamefont
  {Ade} \emph {et~al.}}]{Ade:2015ava}%
  \BibitemOpen
  \bibfield  {author} {\bibinfo {author} {\bibfnamefont {P.~A.~R.}\
  \bibnamefont {Ade}} \emph {et~al.} (\bibinfo {collaboration} {Planck}),\
  }\href {https://doi.org/10.1051/0004-6361/201525836} {\bibfield  {journal}
  {\bibinfo  {journal} {Astron. Astrophys.}\ }\textbf {\bibinfo {volume}
  {594}},\ \bibinfo {pages} {A17} (\bibinfo {year} {2016}{\natexlab{b}})},\
  \Eprint {https://arxiv.org/abs/1502.01592} {arXiv:1502.01592 [astro-ph.CO]}
  \BibitemShut {NoStop}%
\bibitem [{\citenamefont {Akrami}\ \emph
  {et~al.}(2020{\natexlab{b}})\citenamefont {Akrami} \emph
  {et~al.}}]{Akrami:2019izv}%
  \BibitemOpen
  \bibfield  {author} {\bibinfo {author} {\bibfnamefont {Y.}~\bibnamefont
  {Akrami}} \emph {et~al.} (\bibinfo {collaboration} {Planck}),\ }\href
  {https://doi.org/10.1051/0004-6361/201935891} {\bibfield  {journal} {\bibinfo
   {journal} {Astron. Astrophys.}\ }\textbf {\bibinfo {volume} {641}},\
  \bibinfo {pages} {A9} (\bibinfo {year} {2020}{\natexlab{b}})},\ \Eprint
  {https://arxiv.org/abs/1905.05697} {arXiv:1905.05697 [astro-ph.CO]}
  \BibitemShut {NoStop}%
\bibitem [{\citenamefont {Adshead}\ and\ \citenamefont
  {Wyman}(2012{\natexlab{a}})}]{Adshead:2012kp}%
  \BibitemOpen
  \bibfield  {author} {\bibinfo {author} {\bibfnamefont {P.}~\bibnamefont
  {Adshead}}\ and\ \bibinfo {author} {\bibfnamefont {M.}~\bibnamefont
  {Wyman}},\ }\href {https://doi.org/10.1103/PhysRevLett.108.261302} {\bibfield
   {journal} {\bibinfo  {journal} {Phys. Rev. Lett.}\ }\textbf {\bibinfo
  {volume} {108}},\ \bibinfo {pages} {261302} (\bibinfo {year}
  {2012}{\natexlab{a}})},\ \Eprint {https://arxiv.org/abs/1202.2366}
  {arXiv:1202.2366 [hep-th]} \BibitemShut {NoStop}%
\bibitem [{\citenamefont {Maleknejad}\ and\ \citenamefont
  {Sheikh-Jabbari}(2013)}]{Maleknejad:2011jw}%
  \BibitemOpen
  \bibfield  {author} {\bibinfo {author} {\bibfnamefont {A.}~\bibnamefont
  {Maleknejad}}\ and\ \bibinfo {author} {\bibfnamefont {M.~M.}\ \bibnamefont
  {Sheikh-Jabbari}},\ }\href {https://doi.org/10.1016/j.physletb.2013.05.001}
  {\bibfield  {journal} {\bibinfo  {journal} {Phys. Lett. B}\ }\textbf
  {\bibinfo {volume} {723}},\ \bibinfo {pages} {224} (\bibinfo {year}
  {2013})},\ \Eprint {https://arxiv.org/abs/1102.1513} {arXiv:1102.1513
  [hep-ph]} \BibitemShut {NoStop}%
\bibitem [{\citenamefont {Maleknejad}\ and\ \citenamefont
  {Sheikh-Jabbari}(2011)}]{Maleknejad:2011sq}%
  \BibitemOpen
  \bibfield  {author} {\bibinfo {author} {\bibfnamefont {A.}~\bibnamefont
  {Maleknejad}}\ and\ \bibinfo {author} {\bibfnamefont {M.~M.}\ \bibnamefont
  {Sheikh-Jabbari}},\ }\href {https://doi.org/10.1103/PhysRevD.84.043515}
  {\bibfield  {journal} {\bibinfo  {journal} {Phys. Rev. D}\ }\textbf {\bibinfo
  {volume} {84}},\ \bibinfo {pages} {043515} (\bibinfo {year} {2011})},\
  \Eprint {https://arxiv.org/abs/1102.1932} {arXiv:1102.1932 [hep-ph]}
  \BibitemShut {NoStop}%
\bibitem [{\citenamefont {Adshead}\ and\ \citenamefont
  {Wyman}(2012{\natexlab{b}})}]{Adshead:2012qe}%
  \BibitemOpen
  \bibfield  {author} {\bibinfo {author} {\bibfnamefont {P.}~\bibnamefont
  {Adshead}}\ and\ \bibinfo {author} {\bibfnamefont {M.}~\bibnamefont
  {Wyman}},\ }\href {https://doi.org/10.1103/PhysRevD.86.043530} {\bibfield
  {journal} {\bibinfo  {journal} {Phys. Rev. D}\ }\textbf {\bibinfo {volume}
  {86}},\ \bibinfo {pages} {043530} (\bibinfo {year} {2012}{\natexlab{b}})},\
  \Eprint {https://arxiv.org/abs/1203.2264} {arXiv:1203.2264 [hep-th]}
  \BibitemShut {NoStop}%
\bibitem [{\citenamefont {Namba}\ \emph {et~al.}(2013)\citenamefont {Namba},
  \citenamefont {Dimastrogiovanni},\ and\ \citenamefont
  {Peloso}}]{Namba:2013kia}%
  \BibitemOpen
  \bibfield  {author} {\bibinfo {author} {\bibfnamefont {R.}~\bibnamefont
  {Namba}}, \bibinfo {author} {\bibfnamefont {E.}~\bibnamefont
  {Dimastrogiovanni}},\ and\ \bibinfo {author} {\bibfnamefont {M.}~\bibnamefont
  {Peloso}},\ }\href {https://doi.org/10.1088/1475-7516/2013/11/045} {\bibfield
   {journal} {\bibinfo  {journal} {JCAP}\ }\textbf {\bibinfo {volume} {11}},\
  \bibinfo {pages} {045}},\ \Eprint {https://arxiv.org/abs/1308.1366}
  {arXiv:1308.1366 [astro-ph.CO]} \BibitemShut {NoStop}%
\bibitem [{\citenamefont {Maleknejad}\ and\ \citenamefont
  {Erfani}(2014)}]{Maleknejad:2013npa}%
  \BibitemOpen
  \bibfield  {author} {\bibinfo {author} {\bibfnamefont {A.}~\bibnamefont
  {Maleknejad}}\ and\ \bibinfo {author} {\bibfnamefont {E.}~\bibnamefont
  {Erfani}},\ }\href {https://doi.org/10.1088/1475-7516/2014/03/016} {\bibfield
   {journal} {\bibinfo  {journal} {JCAP}\ }\textbf {\bibinfo {volume} {03}},\
  \bibinfo {pages} {016}},\ \Eprint {https://arxiv.org/abs/1311.3361}
  {arXiv:1311.3361 [hep-th]} \BibitemShut {NoStop}%
\bibitem [{\citenamefont {Wolfson}\ \emph {et~al.}(2020)\citenamefont
  {Wolfson}, \citenamefont {Maleknejad},\ and\ \citenamefont
  {Komatsu}}]{Wolfson:2020fqz}%
  \BibitemOpen
  \bibfield  {author} {\bibinfo {author} {\bibfnamefont {I.}~\bibnamefont
  {Wolfson}}, \bibinfo {author} {\bibfnamefont {A.}~\bibnamefont
  {Maleknejad}},\ and\ \bibinfo {author} {\bibfnamefont {E.}~\bibnamefont
  {Komatsu}},\ }\href {https://doi.org/10.1088/1475-7516/2020/09/047}
  {\bibfield  {journal} {\bibinfo  {journal} {JCAP}\ }\textbf {\bibinfo
  {volume} {09}},\ \bibinfo {pages} {047}},\ \Eprint
  {https://arxiv.org/abs/2003.01617} {arXiv:2003.01617 [gr-qc]} \BibitemShut
  {NoStop}%
\bibitem [{\citenamefont {Wolfson}\ \emph {et~al.}(2021)\citenamefont
  {Wolfson}, \citenamefont {Maleknejad}, \citenamefont {Murata}, \citenamefont
  {Komatsu},\ and\ \citenamefont {Kobayashi}}]{Wolfson:2021fya}%
  \BibitemOpen
  \bibfield  {author} {\bibinfo {author} {\bibfnamefont {I.}~\bibnamefont
  {Wolfson}}, \bibinfo {author} {\bibfnamefont {A.}~\bibnamefont {Maleknejad}},
  \bibinfo {author} {\bibfnamefont {T.}~\bibnamefont {Murata}}, \bibinfo
  {author} {\bibfnamefont {E.}~\bibnamefont {Komatsu}},\ and\ \bibinfo {author}
  {\bibfnamefont {T.}~\bibnamefont {Kobayashi}},\ }\href
  {https://doi.org/10.1088/1475-7516/2021/09/031} {\bibfield  {journal}
  {\bibinfo  {journal} {JCAP}\ }\textbf {\bibinfo {volume} {09}},\ \bibinfo
  {pages} {031}},\ \Eprint {https://arxiv.org/abs/2105.06259} {arXiv:2105.06259
  [gr-qc]} \BibitemShut {NoStop}%
\bibitem [{\citenamefont {Dimastrogiovanni}\ and\ \citenamefont
  {Peloso}(2013)}]{Dimastrogiovanni:2012ew}%
  \BibitemOpen
  \bibfield  {author} {\bibinfo {author} {\bibfnamefont {E.}~\bibnamefont
  {Dimastrogiovanni}}\ and\ \bibinfo {author} {\bibfnamefont {M.}~\bibnamefont
  {Peloso}},\ }\href {https://doi.org/10.1103/PhysRevD.87.103501} {\bibfield
  {journal} {\bibinfo  {journal} {Phys. Rev. D}\ }\textbf {\bibinfo {volume}
  {87}},\ \bibinfo {pages} {103501} (\bibinfo {year} {2013})},\ \Eprint
  {https://arxiv.org/abs/1212.5184} {arXiv:1212.5184 [astro-ph.CO]}
  \BibitemShut {NoStop}%
\bibitem [{\citenamefont {Adshead}\ \emph
  {et~al.}(2013{\natexlab{a}})\citenamefont {Adshead}, \citenamefont
  {Martinec},\ and\ \citenamefont {Wyman}}]{Adshead:2013qp}%
  \BibitemOpen
  \bibfield  {author} {\bibinfo {author} {\bibfnamefont {P.}~\bibnamefont
  {Adshead}}, \bibinfo {author} {\bibfnamefont {E.}~\bibnamefont {Martinec}},\
  and\ \bibinfo {author} {\bibfnamefont {M.}~\bibnamefont {Wyman}},\ }\href
  {https://doi.org/10.1103/PhysRevD.88.021302} {\bibfield  {journal} {\bibinfo
  {journal} {Phys. Rev. D}\ }\textbf {\bibinfo {volume} {88}},\ \bibinfo
  {pages} {021302} (\bibinfo {year} {2013}{\natexlab{a}})},\ \Eprint
  {https://arxiv.org/abs/1301.2598} {arXiv:1301.2598 [hep-th]} \BibitemShut
  {NoStop}%
\bibitem [{\citenamefont {Adshead}\ \emph
  {et~al.}(2013{\natexlab{b}})\citenamefont {Adshead}, \citenamefont
  {Martinec},\ and\ \citenamefont {Wyman}}]{Adshead:2013nka}%
  \BibitemOpen
  \bibfield  {author} {\bibinfo {author} {\bibfnamefont {P.}~\bibnamefont
  {Adshead}}, \bibinfo {author} {\bibfnamefont {E.}~\bibnamefont {Martinec}},\
  and\ \bibinfo {author} {\bibfnamefont {M.}~\bibnamefont {Wyman}},\ }\href
  {https://doi.org/10.1007/JHEP09(2013)087} {\bibfield  {journal} {\bibinfo
  {journal} {JHEP}\ }\textbf {\bibinfo {volume} {09}},\ \bibinfo {pages}
  {087}},\ \Eprint {https://arxiv.org/abs/1305.2930} {arXiv:1305.2930 [hep-th]}
  \BibitemShut {NoStop}%
\bibitem [{\citenamefont {Dimastrogiovanni}\ \emph {et~al.}(2017)\citenamefont
  {Dimastrogiovanni}, \citenamefont {Fasiello},\ and\ \citenamefont
  {Fujita}}]{Dimastrogiovanni:2016fuu}%
  \BibitemOpen
  \bibfield  {author} {\bibinfo {author} {\bibfnamefont {E.}~\bibnamefont
  {Dimastrogiovanni}}, \bibinfo {author} {\bibfnamefont {M.}~\bibnamefont
  {Fasiello}},\ and\ \bibinfo {author} {\bibfnamefont {T.}~\bibnamefont
  {Fujita}},\ }\href {https://doi.org/10.1088/1475-7516/2017/01/019} {\bibfield
   {journal} {\bibinfo  {journal} {JCAP}\ }\textbf {\bibinfo {volume} {01}},\
  \bibinfo {pages} {019}},\ \Eprint {https://arxiv.org/abs/1608.04216}
  {arXiv:1608.04216 [astro-ph.CO]} \BibitemShut {NoStop}%
\bibitem [{\citenamefont {Agrawal}\ \emph
  {et~al.}(2018{\natexlab{b}})\citenamefont {Agrawal}, \citenamefont {Fujita},\
  and\ \citenamefont {Komatsu}}]{Agrawal:2017awz}%
  \BibitemOpen
  \bibfield  {author} {\bibinfo {author} {\bibfnamefont {A.}~\bibnamefont
  {Agrawal}}, \bibinfo {author} {\bibfnamefont {T.}~\bibnamefont {Fujita}},\
  and\ \bibinfo {author} {\bibfnamefont {E.}~\bibnamefont {Komatsu}},\ }\href
  {https://doi.org/10.1103/PhysRevD.97.103526} {\bibfield  {journal} {\bibinfo
  {journal} {Phys. Rev. D}\ }\textbf {\bibinfo {volume} {97}},\ \bibinfo
  {pages} {103526} (\bibinfo {year} {2018}{\natexlab{b}})},\ \Eprint
  {https://arxiv.org/abs/1707.03023} {arXiv:1707.03023 [astro-ph.CO]}
  \BibitemShut {NoStop}%
\bibitem [{\citenamefont {Agrawal}\ \emph
  {et~al.}(2018{\natexlab{c}})\citenamefont {Agrawal}, \citenamefont {Fujita},\
  and\ \citenamefont {Komatsu}}]{Agrawal:2018mrg}%
  \BibitemOpen
  \bibfield  {author} {\bibinfo {author} {\bibfnamefont {A.}~\bibnamefont
  {Agrawal}}, \bibinfo {author} {\bibfnamefont {T.}~\bibnamefont {Fujita}},\
  and\ \bibinfo {author} {\bibfnamefont {E.}~\bibnamefont {Komatsu}},\ }\href
  {https://doi.org/10.1088/1475-7516/2018/06/027} {\bibfield  {journal}
  {\bibinfo  {journal} {JCAP}\ }\textbf {\bibinfo {volume} {06}},\ \bibinfo
  {pages} {027}},\ \Eprint {https://arxiv.org/abs/1802.09284} {arXiv:1802.09284
  [astro-ph.CO]} \BibitemShut {NoStop}%
\bibitem [{\citenamefont {Dimastrogiovanni}\ \emph {et~al.}(2018)\citenamefont
  {Dimastrogiovanni}, \citenamefont {Fasiello}, \citenamefont {Hardwick},
  \citenamefont {Assadullahi}, \citenamefont {Koyama},\ and\ \citenamefont
  {Wands}}]{Dimastrogiovanni:2018xnn}%
  \BibitemOpen
  \bibfield  {author} {\bibinfo {author} {\bibfnamefont {E.}~\bibnamefont
  {Dimastrogiovanni}}, \bibinfo {author} {\bibfnamefont {M.}~\bibnamefont
  {Fasiello}}, \bibinfo {author} {\bibfnamefont {R.~J.}\ \bibnamefont
  {Hardwick}}, \bibinfo {author} {\bibfnamefont {H.}~\bibnamefont
  {Assadullahi}}, \bibinfo {author} {\bibfnamefont {K.}~\bibnamefont
  {Koyama}},\ and\ \bibinfo {author} {\bibfnamefont {D.}~\bibnamefont
  {Wands}},\ }\href {https://doi.org/10.1088/1475-7516/2018/11/029} {\bibfield
  {journal} {\bibinfo  {journal} {JCAP}\ }\textbf {\bibinfo {volume} {11}},\
  \bibinfo {pages} {029}},\ \Eprint {https://arxiv.org/abs/1806.05474}
  {arXiv:1806.05474 [astro-ph.CO]} \BibitemShut {NoStop}%
\bibitem [{\citenamefont {Fujita}\ \emph {et~al.}(2019)\citenamefont {Fujita},
  \citenamefont {Namba},\ and\ \citenamefont {Obata}}]{Fujita:2018vmv}%
  \BibitemOpen
  \bibfield  {author} {\bibinfo {author} {\bibfnamefont {T.}~\bibnamefont
  {Fujita}}, \bibinfo {author} {\bibfnamefont {R.}~\bibnamefont {Namba}},\ and\
  \bibinfo {author} {\bibfnamefont {I.}~\bibnamefont {Obata}},\ }\href
  {https://doi.org/10.1088/1475-7516/2019/04/044} {\bibfield  {journal}
  {\bibinfo  {journal} {JCAP}\ }\textbf {\bibinfo {volume} {04}},\ \bibinfo
  {pages} {044}},\ \Eprint {https://arxiv.org/abs/1811.12371} {arXiv:1811.12371
  [astro-ph.CO]} \BibitemShut {NoStop}%
\bibitem [{\citenamefont {Fujita}\ \emph
  {et~al.}(2021{\natexlab{a}})\citenamefont {Fujita}, \citenamefont {Murai},
  \citenamefont {Obata},\ and\ \citenamefont {Shiraishi}}]{Fujita:2021flu}%
  \BibitemOpen
  \bibfield  {author} {\bibinfo {author} {\bibfnamefont {T.}~\bibnamefont
  {Fujita}}, \bibinfo {author} {\bibfnamefont {K.}~\bibnamefont {Murai}},
  \bibinfo {author} {\bibfnamefont {I.}~\bibnamefont {Obata}},\ and\ \bibinfo
  {author} {\bibfnamefont {M.}~\bibnamefont {Shiraishi}}\ }\href
  {https://doi.org/10.1088/1475-7516/2022/01/007}
  {10.1088/1475-7516/2022/01/007} (\bibinfo {year} {2021}{\natexlab{a}}),\
  \Eprint {https://arxiv.org/abs/2109.06457} {arXiv:2109.06457 [astro-ph.CO]}
  \BibitemShut {NoStop}%
\bibitem [{\citenamefont {Ishiwata}\ \emph {et~al.}(2022)\citenamefont
  {Ishiwata}, \citenamefont {Komatsu},\ and\ \citenamefont
  {Obata}}]{Ishiwata:2021yne}%
  \BibitemOpen
  \bibfield  {author} {\bibinfo {author} {\bibfnamefont {K.}~\bibnamefont
  {Ishiwata}}, \bibinfo {author} {\bibfnamefont {E.}~\bibnamefont {Komatsu}},\
  and\ \bibinfo {author} {\bibfnamefont {I.}~\bibnamefont {Obata}},\ }\href
  {https://doi.org/10.1088/1475-7516/2022/03/010} {\bibfield  {journal}
  {\bibinfo  {journal} {JCAP}\ }\textbf {\bibinfo {volume} {03}}\bibfield
  {number} {\bibinfo  {number} { (03)},\ \bibinfo {pages} {010}},\ }\Eprint
  {https://arxiv.org/abs/2111.14429} {arXiv:2111.14429 [hep-ph]} \BibitemShut
  {NoStop}%
\bibitem [{\citenamefont {Fujita}\ \emph
  {et~al.}(2021{\natexlab{b}})\citenamefont {Fujita}, \citenamefont
  {Nakatsuka}, \citenamefont {Mukaida},\ and\ \citenamefont
  {Murai}}]{Fujita:2021eue}%
  \BibitemOpen
  \bibfield  {author} {\bibinfo {author} {\bibfnamefont {T.}~\bibnamefont
  {Fujita}}, \bibinfo {author} {\bibfnamefont {H.}~\bibnamefont {Nakatsuka}},
  \bibinfo {author} {\bibfnamefont {K.}~\bibnamefont {Mukaida}},\ and\ \bibinfo
  {author} {\bibfnamefont {K.}~\bibnamefont {Murai}},\ }\href@noop {} {\
  (\bibinfo {year} {2021}{\natexlab{b}})},\ \Eprint
  {https://arxiv.org/abs/2110.03228} {arXiv:2110.03228 [hep-ph]} \BibitemShut
  {NoStop}%
\bibitem [{\citenamefont {Domcke}\ \emph {et~al.}(2019)\citenamefont {Domcke},
  \citenamefont {Mares}, \citenamefont {Muia},\ and\ \citenamefont
  {Pieroni}}]{Domcke:2018rvv}%
  \BibitemOpen
  \bibfield  {author} {\bibinfo {author} {\bibfnamefont {V.}~\bibnamefont
  {Domcke}}, \bibinfo {author} {\bibfnamefont {B.}~\bibnamefont {Mares}},
  \bibinfo {author} {\bibfnamefont {F.}~\bibnamefont {Muia}},\ and\ \bibinfo
  {author} {\bibfnamefont {M.}~\bibnamefont {Pieroni}},\ }\href
  {https://doi.org/10.1088/1475-7516/2019/04/034} {\bibfield  {journal}
  {\bibinfo  {journal} {JCAP}\ }\textbf {\bibinfo {volume} {04}},\ \bibinfo
  {pages} {034}},\ \Eprint {https://arxiv.org/abs/1807.03358} {arXiv:1807.03358
  [hep-ph]} \BibitemShut {NoStop}%
\bibitem [{\citenamefont {Ramond}(2010)}]{ramond_2010}%
  \BibitemOpen
  \bibfield  {author} {\bibinfo {author} {\bibfnamefont {P.}~\bibnamefont
  {Ramond}},\ }\href {https://doi.org/10.1017/CBO9780511781865} {\emph
  {\bibinfo {title} {{Group Theory: A Physicist's Survey}}}}\ (\bibinfo
  {publisher} {Cambridge University Press},\ \bibinfo {year}
  {2010})\BibitemShut {NoStop}%
\end{thebibliography}%

\end{document}